\documentclass[prl,twocolumn,altaffilletter,nolongbibliography,numerical,flushbottom,secnumarabic,superscriptaddress,floatfix]{revtex4-2}
\usepackage[utf8]{inputenc}
\usepackage[american]{babel}
\usepackage[T1]{fontenc}
\usepackage[pdftex]{graphicx}  
\usepackage[usenames,dvipsnames,svgnames]{xcolor}
 
\usepackage{dcolumn}
\usepackage{comment}
\usepackage{scalerel}
\usepackage{lipsum}
\usepackage{footnote}
\usepackage{amsmath}
\usepackage{amsfonts}
\usepackage{amssymb}
\usepackage{color}
\usepackage[colorlinks,bookmarks=false,citecolor=NavyBlue,linkcolor=Red,urlcolor=blue]{hyperref}
\usepackage[export]{adjustbox}
\usepackage{amsthm}
\usepackage{simplewick}
\usepackage{bbold}
\usepackage{MnSymbol}
\usepackage{physics}
\usepackage{wasysym}
\usepackage[toc]{appendix}
\usepackage{mathrsfs}  
\usepackage{braket}
\usepackage{graphics}
\usepackage{todonotes}
\usepackage{multirow}
\usepackage{algorithm}
\usepackage{algpseudocode}
\usepackage{dsfont}
\usepackage{mdframed}

\newcommand{\Pur}{{\text{P}}} 
\newcommand{\Wg}{{\text{Wg}}} 
\newcommand{\Order}{\mathcal{O}} 
\newcommand{\Cliff}{\mathcal{C}} 

\newcommand{\Id}{\mathbb{1}}
\newcommand{\PauliX}{{\text{X}}}
\newcommand{\PauliY}{{\text{Y}}}
\newcommand{\PauliZ}{{\text{Z}}}

\newcommand{\muCl}{{\mu_{\Cliff}}} 
\newcommand{\muH}{{\mu_{\text{H}}}} 
\newcommand{\murmps}{{\mu_{\chi}}} 

\newcommand{\Ex}{\mathbb{E}} 

\begin{document}

\newcommand{\titleinfo}{Quantum State Designs with Clifford Enhanced Matrix Product States}
\title{\titleinfo}

\author{Guglielmo Lami}
\affiliation{Laboratoire de Physique Th\'eorique et Mod\'elisation, CNRS UMR 8089,
CY Cergy Paris Universit\'e, 95302 Cergy-Pontoise Cedex, France}

\author{Tobias Haug}
\affiliation{Quantum Research Center, Technology Innovation Institute, Abu Dhabi, UAE}

\author{Jacopo De Nardis}
\affiliation{Laboratoire de Physique Th\'eorique et Mod\'elisation, CNRS UMR 8089,
CY Cergy Paris Universit\'e, 95302 Cergy-Pontoise Cedex, France}

\begin{abstract}
Nonstabilizerness, or `magic', is a critical quantum resource that, together with entanglement, characterizes the non-classical complexity of quantum states. Here, we address the problem of quantifying the average nonstabilizerness of random Matrix Product States (RMPS). RMPS represent a generalization of random product states featuring bounded entanglement that scales logarithmically with the bond dimension $\chi$. We demonstrate that the Stabilizer Rényi Entropies converges to that of Haar random states as $N/\chi^\alpha$, where $N$ is the system size and $\alpha$ are integer exponents. This indicates that MPS with a modest bond dimension are as magical as generic states. Subsequently, we introduce the ensemble of Clifford enhanced Matrix Product States ($\mathcal{C}$MPS), built by the action of Clifford unitaries on RMPS. Leveraging our previous result, we show that $\mathcal{C}$MPS can approximate a quantum state $4$-designs with arbitrary accuracy. Specifically, for a constant $N$, $\mathcal{C}$MPS become close to $4$-designs with a scaling as $\chi^{-2}$. Our findings indicate that combining Clifford unitaries with polynomially complex tensor network states can generate highly non-trivial quantum states.
\end{abstract}

\maketitle

\section{Introduction} 
Quantum states of many interacting particles (or qubits) are very complex, since they typically require exponential resources to be simulated~\cite{Feynman1982, RevModPhys.71.1253, Daley_2022,Xu_2023,Pashayan_2020} or learned~\cite{Anshu_2023}. Nevertheless, at least two broad categories of quantum states that can be simulated classically are known. 

Firstly, states exhibiting limited amount of quantum correlations between their constituencies, i.e.\ low entanglement, can be simulated efficiently by means of Tensor Networks~\cite{Vidal_2004,Schollwock_2011,Silvi_2019,Biamonte_2020,Cirac_2021,Stoudenmire_2024}. Their one-dimensional version, Matrix Product States (MPS) involves contracting $N$ matrices of size $\chi$, where $\chi$ represents the bond dimension and $N$ the number of qubits. Entanglement scales as $\log \chi$, ensuring efficiency as long as entanglement remains bounded~\cite{Hastings_2007}.

However, while entanglement is a crucial quantum resource, it is not sufficient for classical hardness. Stabilizer states, formed by the Clifford group acting on computational basis states, may exhibit high entanglement but remain simulatable and learnable with polynomial complexity~\cite{Nielsen_chuang_2010,Gottesman_1997, Gottesman_1998_1, Gottesman_1998_2,Gross_2016,Kueng_2015,Grewal_2023_2,Beverland_2020,Jiang_2023,Montanaro_2017,Iaconis,Oliviero_2022_2,Gullans_2023}.
The Clifford group entails unitary operations mapping Pauli operators into Pauli operators~\cite{Nielsen_chuang_2010}. 
Generic quantum states can be formed by combining Clifford operations and non-Clifford resource states, where the latter are regarded as a key quantum resource for fault-tolerant quantum computation~\cite{Nielsen_chuang_2010,Veitch_2014}. The resource theory of nonstabilizerness (aka quantum magic) has been introduced to quantify the extent to which a state deviates from being constructed solely with Cliffords~\cite{Bravyi_2005,Veitch_2014,Leone_2022,Winter_2022}. States with sufficiently low magic can be simulated using the tableau formalism~\cite{Nielsen_chuang_2010,Gottesman_1997, Gottesman_1998_1, Gottesman_1998_2,Aaronson_2004,Yoganathan_2019,Gu_2024,Pashayan_2022}. However, simulations (such as sampling in the computational basis) can become inefficient when Cliffords interact with a large number of magic resource states~\cite{Yoganathan_2019,Bejan_2023}. States obtained with random Cliffords applied to magical resource have attracted considerable interest, and average values of physical quantities such as multi-fractal flatness~\cite{Turkeshi_2023_1}, out-of-time-order correlators~\cite{Leone_2021,Haug_2023_3} and entanglement spectrum flatness~\cite{Tirrito_2023} are directly linked to the nonstabilizerness of the resource state.

In this paper, we focus on a broader class of magical resource states, namely Matrix Product States (MPS). Indeed, identifying effective methods for combining MPS with Clifford circuits and characterizing the resulting states presents an intriguing and largely unexplored prospect.  Motivated by the quest for new hybrid quantum-classical simulation techniques, here we analyze the expressivity of a Clifford enhanced Matrix Product States $|\psi \rangle =  \mathcal{U}_c | \psi_\chi \rangle $ where $\mathcal{U}_c$ is a Clifford unitary and $| \psi_\chi \rangle $ an MPS state with bond dimension $\chi$. Given that the action of $\mathcal{U}_c$ can arbitrarily increase the amount of entanglement of the MPS state $| \psi_\chi \rangle$, it is legit to postulate that states of the form $|\psi \rangle$ can approximate a large class of generic quantum states. 

\begin{figure}[t!]
\includegraphics[width=1.02\linewidth]{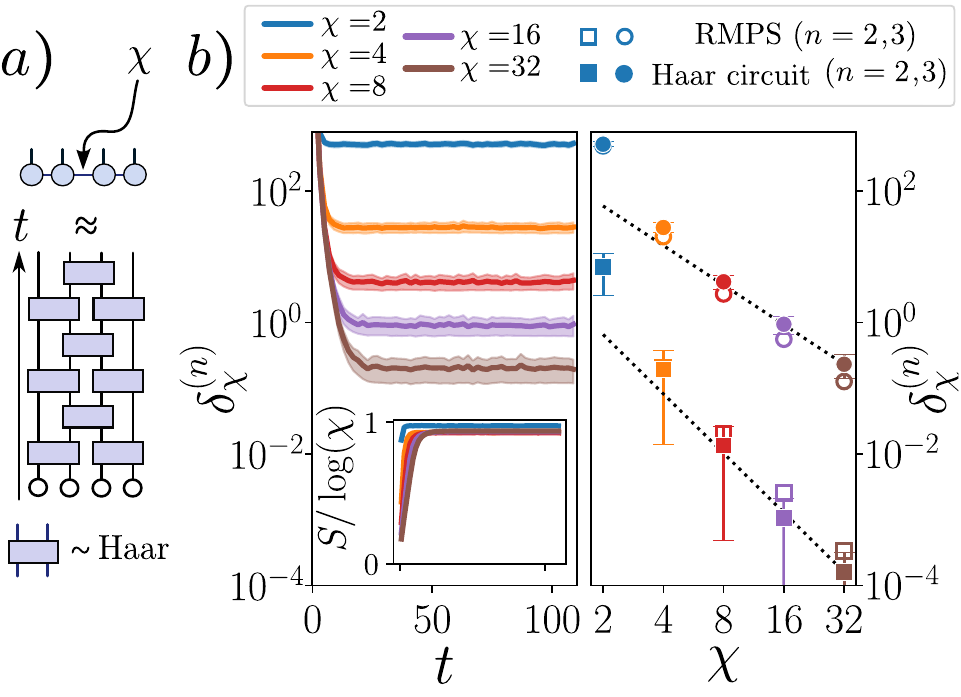}
\caption{$a)$ A random circuit comprising two local random unitary is considered. The initial state evolves as an MPS, with the bond dimension growing up to $\chi$.
$b)$ Left: convergence of the magic with the MPS bond dimension ($N=22$ qubits, $500$ trajectories). Deviation of the $n$-SRE ($n=2$) from the averaged Haar value, $\delta_{\chi}^{(n)}$, is plotted over discrete circuit time $t$. (Inset: maximum entanglement entropy $S$ of the evolved MPS). Right: $\delta_{\chi}^{(n)}$ ($n=2,3$) for the final state compared with exact RMPS average obtained for the same size and $\chi$. Black lines represents power laws $\chi^{-2}$ and $\chi^{-3}$.} 
\label{fig:1}
\end{figure}

\section{Main results}
We focus on typical states resulting from the action of a Clifford unitary on a generic MPS. To characterize such a ensemble, we employ the formalism of random Matrix Product States (RMPS)~\cite{Garnerone_2010_1, Garnerone_2010_2,Haag_2023,Haferkamp_2021,Lancien2021,Chen_2022}. RMPS can be taken as generalisation of random product states with finite entanglement scaling as $\log \chi$, where $\chi$ is the bond dimension. First, we show that for RMPS with open boundary conditions, the Stabiliser Rényi Entropies (SRE)~\cite{Leone_2022}, a widely used measure of nonstabilizerness, approach their average value for random Haar-distributed states with deviations $\Order(N/\chi^2)$ for Rényi index $n=2$, and $\Order(N/\chi^3)$ for Rényi index $n=3$. As illustrated in Fig.~\ref{fig:1}, this behavior can be visualized by evolving an MPS under a random Haar circuit \cite{Fisher2023} truncated to bond dimension $\chi$. Entanglement entropy $S$ rapidly reaches its maximum value $\log \chi$ (see inset). Similarly, the SRE approaches a $\chi$-dependent value, converging polynomially quickly to the Haar state values. We then investigate how the ensemble generated by applying random Clifford operators to RMPS ($\Cliff$MPS) approximates Haar states. For this purpose, we adopt the concept of $k$-design, a property of state ensembles reproducing Haar averages up to the $k$-th moment. Stabilizer states form  exact $3$-designs~\cite{Kueng_2015}, approximate $4$-designs, and are far from $6$-designs~\cite{gross2021schur}. Although the deviation from $4$-design decreases exponentially with the number $N$ of qubits~\cite{Damanik_2018,gross2021schur}, for fixed $N$ this may not be sufficiently small in practice. Indeed, stabilizers do not reproduce important universal features of Haar states, having for instance exponentially larger purity fluctuations~\cite{Leone_2021}. Here, we show that for $\Cliff$MPS the deviation from $4$-design can be arbitrarily reduced by increasing $\chi$, with deviations that scale as $N \chi^{-2}$. 
Given the complexity of computing $k$-folded Clifford channels for $k>4$, we cannot predict higher moments, yet achieving a $4$-design with arbitrary accuracy is significant. Indeed, we prove that the inherent simplicity of MPS and Cliffords, when combined, can produce highly non-trivial quantum states, featuring the same entanglement, purity fluctuations, and magic as typical quantum states. \\ 

\section{Preliminaries}  We consider a system consisting of $N$ qubits. The Hilbert space, of size $d=2^N$, is locally spanned by the basis states $\ket{s_i} \in \{\ket{0}, \ket{1}\}$ ($i=1,2 ...N$). $\mathcal{U}(d)$ is the corresponding group of unitary operators. The Haar measure $\muH$ is the unique left/right invariant measure over $\mathcal{U}(d)$~\cite{mele_2024}. Expectation values over Haar are denoted as $\Ex_{\mathcal{U} \sim \muH}[\dots] \equiv \int d \muH \, (\dots )$. For any ensemble of unitaries $\mathcal{X}$, we can define the uniform probability distribution over it $\mu_{\mathcal{X}}$. Expectation values can be computed as 
$\Ex_{\mathcal{U} \sim \mu_{\mathcal{X}}}[\dots] \equiv 1/|\mathcal{X}| \cdot \sum_{\mathcal{U} \in \mathcal{X}} (\dots )$. An ensemble of unitaries that satisfies
\begin{equation}
\Ex_{\mathcal{U} \sim \mu_{\mathcal{X}}}[(\mathcal{U}^{\dag})^{\otimes k} O (\mathcal{U})^{\otimes k}] = \Ex_{\mathcal{U} \sim \muH}[(\mathcal{U}^{\dag})^{\otimes k} O (\mathcal{U})^{\otimes k}]        
\end{equation}
for any operator $O$ acting on $k$ replicas of the system is called $k$-design. Intuitively, $k$-designs are distributions of unitaries that replicate the moments of the Haar measure up to the $k$-th order. Similarly, if now $\mathcal{X}$ represent a set of states $\{ \ket{\psi} \}$, one can introduce the notion of quantum state $k$-design. This refers to an ensemble for which 
\begin{equation}
\Ex_{\psi \sim \mu_{\mathcal{X}}}[\big( \ket{\psi} \bra{\psi} \big)^{\otimes k}] = \Ex_{\psi \sim \muH}[ \big( \ket{\psi} \bra{\psi} \big)^{\otimes k}] \, .    
\end{equation}
If $\mathcal{X}$ is a $k$-unitary design then the ensemble of states $\mathcal{X} \ket{\pmb{0}}$, with $\ket{\pmb{0}} \equiv \bigotimes_{j=1}^{N} \ket{0}$, is a $k$-design. The frame potential of $\mathcal{X}$ is defined as~\cite{Gross_2007, mele_2024, Ippoliti_2022, Kueng_2015} 
\begin{align*}
       \mathcal{F}^{(k)}_{\mathcal{X}} &= \Ex_{\psi, \psi' \sim \mu_{\mathcal{X}}}[ |\braket{\psi'|\psi}|^{2k}] = \\ &=\Ex_{\psi, \psi'\sim \mu_{\mathcal{X}}} \left[ \Tr[(\ket{\psi'}\bra{\psi'})^{\otimes k} (\ket{\psi}\bra{\psi})^{\otimes k}] \right] \,. 
\end{align*}
It can be proven that $\mathcal{F}^{(k)}_{\mathcal{X}} \geq \mathcal{F}^{(k)}_{\text{H}}$, where $\mathcal{F}^{(k)}_{\text{H}}= \binom{d + k - 1}{k}^{-1}$ is the frame potential of the Haar distribution and the equality holds only if $\mathcal{X}$ is a quantum $k$-design. 

We denote the Pauli operators as $\{\sigma^{\alpha}\}_{\alpha=0}^{3}=\{ \Id, \PauliX, \PauliY, \PauliZ \}$, and with $\pmb{\sigma} = \bigotimes_{j=1}^{N} \sigma_{j}$ a generic tensor product of them, i.e.\ a Pauli string. Pauli strings $\tilde{\mathcal{P}}_N = \{ \pmb{\sigma} \}_{\pmb{\sigma}}$, with appropriate global phases constitute the Pauli group $\mathcal{P}_N = (\pm 1, \pm i) \tilde{\mathcal{P}}_N$. The Clifford group $\Cliff(d)$ consists of $N-$qubits unitaries $\mathcal{U}_c$ preserving the Pauli group under conjugation, i.e.\ $\Cliff(d) = \{ \mathcal{U}_c \in \mathcal{U}(d) \,  \text{ s.t. } \, \mathcal{U}_c^{\dag} \mathcal{P}_N \mathcal{U}_c=\mathcal{P}_N \}$. $\muCl$ represents the uniform probability distribution over the Clifford group, as defined before. Hadamard $H$, phase $S$, and CNOT gates collectively generate the complete group~\cite{Nielsen_chuang_2010}. To approximate any $N$-qubit unitary and achieve universality, an additional gate such as the single-qubit $T$ gate $T = \text{diag}(1, e^{i \pi/4})$ is required~\cite{Nielsen_chuang_2010}. States obtained by applying Cliffords to $\ket{\pmb{0}}$ are known as stabilizer states, and its set is denoted as $\text{STAB}=\{\mathcal{U}_c \ket{\pmb{0}}\}_{\mathcal{U}_c \in \Cliff(d)}$.
There exist classical efficient algorithms to simulate stabilizer states, thanks to a tableau algorithm (Gottesman-Knill theorem)~\cite{Nielsen_chuang_2010, Gottesman_1997, Gottesman_1998_1, Gottesman_1998_2}. \\

\section{Clifford and Haar averages}  
The Clifford group $\Cliff(d)$ constitutes a $3$-unitary design~\cite{Webb_2016}, and therefore $\Ex_{\mathcal{U}_c \sim \muCl}[(\mathcal{U}_c^{\dag})^{\otimes k} O (\mathcal{U}_c)^{\otimes k}] = \Ex_{\mathcal{U} \sim \muH}[(\mathcal{U}^{\dag})^{\otimes k} O (\mathcal{U})^{\otimes k}]$ for $k \leq 3$. Consequently, the set of stabilizer states  constitutes a quantum $3$-design~\cite{Kueng_2015}. The $k-$fold Haar channel $\Ex_{\mathcal{U} \sim \muH}[(\mathcal{U}^{\dag})^{\otimes k} O (\mathcal{U})^{\otimes k}]$ can be computed thanks to the Weingarten calculus~\cite{mele_2024,Kostenberger_2021}, giving
\begin{equation}\label{eq:haark}
    \Ex_{\mathcal{U} \sim \muH}[(\mathcal{U}^{\dag})^{\otimes k} O (\mathcal{U})^{\otimes k}] = \sum_{\sigma, \pi \in S_k} \Wg(\sigma^{-1} \pi, d) \Tr[O T_{\sigma}] T_{\pi} \, ,
\end{equation}
where $T_{\sigma},T_{\pi}$ are unitary representations of the permutations $\sigma, \pi \in S_k$ acting on $k$ replicas of the system, and $\Wg$ are the Weingarten functions. For any normalized state $\ket{\psi}$ and $O=(\ket{\psi} \bra{\psi})^{\otimes k}$, Eq.~\ref{eq:haark} simplifies to $\Ex_{\mathcal{U} \sim \muH}[(\mathcal{U}^{\dag} \ket{\psi} \bra{\psi} \mathcal{U})^{\otimes k}] = P^{(k)}_{\text{symm}}/ \Tr[P^{(k)}_{\text{symm}}]$,
with $P^{(k)}_{\text{symm}} = \sum_{\pi \in S_k} T_{\pi}/k!$
the projector onto the symmetric subspace of the permutation group $S_k$ and $\Tr[P^{(k)}_{\text{symm}}] = \binom{k + d -1}{k} \, .$
As a significant application, we mention the calculation of the purity of a reduced density matrix $\Pur_A(\ket{\psi})= \Tr_A[\rho_A^2]$, $\rho_A = \Tr_{B}[\rho]$ corresponding to a generic system bipartition $A,B$ averaged over the ensemble of pure stabilizer states $\rho=\ket{\psi}\bra{\psi}$, $\ket{\psi} \in $ STAB. First, we rewrite the purity as $\Pur_A(\ket{\psi})= \Tr[\rho^{\otimes 2} T^{(A)}_{21}]$, where $T^{(A)}_{21}$ is the permutation (swap) operator exchanging replicas $1$ and $2$ of subsystem $A$ (while acting as the identity on $B$). Therefore
\begin{equation}
\mathbb{E}_{\psi \sim \text{STAB}} \left[ \Pur_A(\ket{\psi}) \right] = \Tr\left[ \mathbb{E}_{\mathcal{U}_c \sim \muCl}[(\mathcal{U}_c^{\dag} \ket{\pmb{0}} \bra{\pmb{0}} \mathcal{U}_c)^{\otimes 2}] \,  T^{(A)}_{21}\right] \, ,
\end{equation}
A straightforward computation therefore gives
\begin{equation}\label{eq:averagepurity}
\mathbb{E}_{\psi \sim \text{STAB}} \left[ \Pur_A(\ket{\psi}) \right] = \frac{d_A + d_B}{d_A d_B + 1} = \frac{2 \sqrt{d}}{d + 1} \sim \frac{2}{\sqrt{d}} \, ,
\end{equation}
where second equalities holds for $d_A=d_B=d^{1/2}$.
Since the Rényi-2 entanglement entropy is $S_2(\rho_A)= - \log \Pur_A(\ket{\psi})$, 
and $\mathbb{E}\left[ S_2(\rho_A) \right] \geq - \log \mathbb{E} \left[\Pur_A(\ket{\psi}\right])$,
this result shows that stabilizer states have in average the same entanglement of random states, i.e.\ close to the maximum value (which is given by the maximally mixed state $\rho_A = \Id/\sqrt{d}$). \\

The $4-$fold Clifford channel is also known~\cite{Gross_2016}. In particular
\begin{equation}\label{eq:cliff4moment}
\mathbb{E}_{\mathcal{U}_c \sim \muCl} \left[ \big(\mathcal{U}_c^{\dag} \ket{\psi}\bra{\psi} \mathcal{U}_c \big)^{\otimes 4}
\right] = \alpha_{\psi} Q P^{(4)}_{\text{symm}} + \beta_{\psi} P^{(4)}_{\text{symm}}
\end{equation}
where $\displaystyle Q = d^{-2} \sum_{\pmb{\sigma} \in \tilde{\mathcal{P}}_N} \pmb{\sigma}^{\otimes 4}$ and the factors $\alpha_{\psi}$ and $\beta_{\psi}$ are
\begin{equation}\label{eq:cliff4momentcoeff}
    \alpha_{\psi} = \frac{6d(d+3) ||\Pi_{\psi}||_2^2-24}{(d^2-1)(d+2)(d+4)} \quad
    \beta_{\psi} = \frac{24(1-||\Pi_{\psi}||_2^2)}{(d^2-1)(d+2)(d+4)} \, .
\end{equation}
with $\Pi_{\psi}(\pmb{\sigma}) = d^{-1} \braket{\psi|\pmb{\sigma}|\psi}^2$ being a vector of length $d^2$. Note that $\Pi_{\psi}(\pmb{\sigma})$ sums to 1, since 
$d^{-1} \sum_{\pmb{\sigma} \in \tilde{\mathcal{P}}_N} \braket{\psi|\pmb{\sigma}|\psi}^2 = \Tr[\rho^2] =1$, so $\Pi_{\psi}$ is a probability distribution~\cite{Leone_2022}, sometimes referred to as a characteristic function~\cite{Gross_2016}. \\

Using this result, it is possible to compute the fluctuations of purity for stabilizer states~\cite{Leone_2021}. These are defined as:
\begin{equation}\label{eq:purfluc}
\Delta^2 \Pur_A(\ket{\psi}) = \mathbb{E} \left[ \Pur_A(\ket{\psi})^2 \right] - \mathbb{E} \left[ \Pur_A(\ket{\psi}) \right]^2 \, ,
\end{equation}
where the second term, as shown before, is the same for Clifford and Haar random states, while the first term can be computed via Eq.\ref{eq:cliff4moment} (see Appendices for more details). One gets (with $d_A=d_B=d^{1/2}$)~\cite{Leone_2021}
\begin{equation}\label{eq:purfluccliff} 
\Delta^2 \Pur_A(\ket{\psi}) \big\rvert_{\text{STAB}} = \frac{(d-1)^2}{(d+1)^2(d+2)} \sim \Order(d^{-1}) \, ,
\end{equation}
whereas the result for Haar states is
\begin{equation}\label{eq:purfluchaar} 
\Delta^2 \Pur_A(\ket{\psi}) \big\rvert_{\text{H}} = \frac{2(d-1)^2}{(d+1)^2(d+2)(d+3)} \sim \Order(d^{-2})
\end{equation}
Therefore, stabilizer states have exponentially larger purity fluctuations compared to Haar states~\cite{Leone_2021}. \\

\section{Nonstabilizerness}
Nonstabilizerness (magic) is regarded as a veritable resource for quantum systems and computation, both theoretically and practically~\cite{Nielsen_chuang_2010,Veitch_2014,Nielsen_chuang_2010,Gottesman_1997, Gottesman_1998_1, Gottesman_1998_2,Aaronson_2004,Yoganathan_2019,Gu_2024,Pashayan_2022}. In a nutshell, nonstabilizerness is the resource required for quantum states to be unattainable through Clifford circuits (plus Pauli measurements). The latter are regarded as easy to implement free operations, while non-Clifford gates are resources. Several measures of nonstabilizerness have been proposed~\cite{Winter_2022, Howard_2017, Haug_2023}. 

Here, we utilize the Stabilizer Rényi Entropies (SRE), originally introduced in Ref.~\cite{Leone_2022}, which have gained significant attention for their potential in analytical~\cite{Oliviero_2022_1,Passarelli_2024,haug2024probing} and numerical~\cite{Lami_2023_2,Lami_2024,Haug_2023_1,Haug_2023_2,Tarabunga_2024} evaluation, as well as their applicability to experimental measurements~\cite{Gullans_2023,Haug_2023_3}, and their relevance to many-body quantum systems~\cite{Rattacaso_2023,Tarabunga_2023_1, Fux_2023,Bejan_2023,Lami_2024,Turkeshi_2023_2,Gu_2024}.
SRE, indexed by a $n$-Rényi index, are the entropies of the characteristic probability distribution $\Pi_{\psi}(\pmb{\sigma})$ (excluding an additive constant). If we introduce the $(n-1)-$moment of $\Pi_{\psi}$ ($n>1$) as 
\begin{equation}\label{eq:linsre}
    m_n(\ket{\psi}) = ||\Pi_{\psi}||_n^n = d^{-n}\sum_{\pmb{\sigma}}  \braket{\psi|\pmb{\sigma}|\psi}^{2n} \, ,
\end{equation}
the $n-$SRE is given by $\mathcal{M}_n(\ket{\psi}) = (1-n)^{-1} \log m_n(\ket{\psi}) - \log d$, while its linearized version $\mathcal{M}_n^{\text{lin}}(\ket{\psi}) = 1 - d^{n-1} m_n$ is essentially $m_n$, apart from irrelevant constants. SREs exhibit the following properties~\cite{Leone_2022,Leone_2024}, accordingly being a good measure of nonstabilizerness: i) $\mathcal{M}_n \geq 0$ ($m_n \leq d^{1-n}$), with equality holding iff $\ket{\psi} \in $ STAB; ii) SREs are invariant under Clifford unitaries; iii) they are additive. Recently, for pure states the monotonicity of SRE for $n \geq 2$ has been rigorously established~\cite{Leone_2024}, and it has been shown that  $\mathcal{M}_n^{\text{lin}}$ serves as a strong monotone for the same range of $n$. On the other hand, a violation of monotonicity for $0 \leq n < 2$ has been reported in systems undergoing measurements in the computational basis~\cite{Haug_2023_2}. Finally, SREs provide useful bounds for other measures of magic~\cite{Leone_2024}.
The evaluation of SREs for MPS has been explored both through a replica approach~\cite{Haug_2023_1,Haug_2023_2,Tarabunga_2024} and through efficient estimation obtained with a sampling in the Pauli basis~\cite{Lami_2023_2,Lami_2024}. However, the question of how SREs converge with respect to the bond dimension $\chi$ has not been addressed until now. Here, we offer a solution leveraging the RMPS framework. \\

\section{Random Matrix Product States (RMPS)}
MPS~\cite{Schollwock_2011,Silvi_2019,Biamonte_2020} are a class of quantum many-body states characterized by bounded entanglement entropy. MPS are defined as 
\begin{equation}
  \ket{\phi} = \sum_{s_1,s_2,\dots,s_N} \mathbb{A}^{s_1}_{1}\mathbb{A}^{s_2}_{2} \cdots \mathbb{A}^{s_N}_{N} |s_1,s_2,\dots,s_N \rangle  \, ,
\end{equation}
with $\mathbb{A}^{s_{j}}_{j}$ being $\chi_{j-1}\times\chi_{j}$ matrices, except at the left (right) boundary where $\mathbb{A}^{s_{1}}_{1}$ ($\mathbb{A}^{s_{N}}_{N}$) is a $1\times\chi_1$ ($\chi_{N-1}\times1$) row (column) vector. MPS can be graphically represented as~\cite{Schollwock_2011,Silvi_2019,Biamonte_2020} 
\begin{equation}
\includegraphics[width=0.6\linewidth, valign=c]{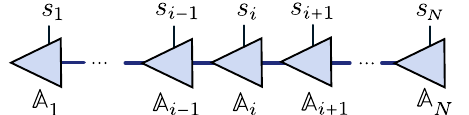}
\end{equation}
Without loss of generality, the tensors $\mathbb{A}$ can be assumed to be isometries, satisfying the right-normalization condition $(\mathbb{A}_i^{s_i})_{\alpha \beta} (\mathbb{A}_i^{s_i,*})_{\alpha' \beta} = \delta_{\alpha \alpha'}$, where indices $\alpha, \alpha', \beta$ run in the auxiliary space, which has dimension $\chi$ (bond dimension)~\cite{Schollwock_2011}. Isometries $\mathbb{A}_i$ can be embedded into larger unitary matrices $U^{(i)}$ of size $q=2 \chi$. The embedding is such that $(\mathbb{A}_i^{s_i})_{\alpha \beta}=\braket{0 \alpha|U^{(i)}|s_i \beta}$, i.e.\
\begin{equation}
\includegraphics[width=0.4\linewidth, valign=c]{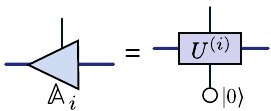} \, .
\end{equation}
As a result, an MPS with maximum bond dimension $\chi$ (assuming $\chi$ is a power of $2$) is equivalent to a quantum circuit comprising sequential unitaries $U^{(i)}$ acting on, at most, $\log_2 \chi + 1$ qubits initialized to $\ket{0}$~\cite{Schon_2005, Lin_2021, Rudolph_2022, Lami_2023_1}, i.e.\
\begin{equation}\label{eq:mps_unitaries_bulk}
\includegraphics[width=0.55\linewidth, valign=c]{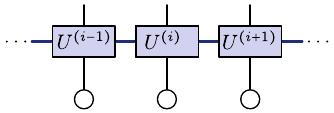}
\end{equation}
At this point, one can introduce a probability measure $\murmps$ for RMPS by requiring the generating unitaries to be Haar random, i.e.\ $U^{(i)} \sim \muH$~\cite{Garnerone_2010_1, Garnerone_2010_2,Haag_2023,Haferkamp_2021}. It should be noted, however, that using $N$ unitaries of the same dimension as in the bulk typically yields unnormalized states, with normalization only restored for large $N$~\cite{Haferkamp_2021}. When handling finite $N$, implementing suitable boundary conditions, with smaller sizes for unitaries at the edges, might be beneficial (see Sec.~\ref{sec:rmpsobc}).

The $k-$fold RMPS channel $\mathbb{E}_{\mathcal{\psi} \sim \murmps}[ \big( \ket{\psi} \bra{\psi} \big)^{\otimes k}
]$ can be constructed by tensorizing $k$ replicas of the Haar matrices $U$, $U^{*}$ (the latter indicated by darker shapes in the following pictures) and applying Eq.~\ref{eq:haark}. The result is~\cite{Haag_2023}
\begin{equation}
\includegraphics[width=0.73\linewidth, valign=c]{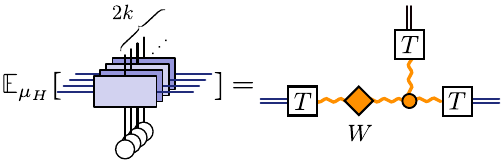}
\end{equation}
Here, we used double lines to represent physical or auxiliary indices tensorized over the replicas, and wavy orange lines to represent permutations indices $\pi \in S_k$. The tensor $T$ contains all permutation operators $T_{\pi}$, folded as vectors~\cite{Haag_2023}. $W$ is the Weingarten matrix $W_{\sigma \pi} = \Wg(\sigma^{-1} \pi, q)$~\cite{Kostenberger_2021}. We obtain therefore
\begin{equation}\label{eq:mps_average_k_bulk}
\includegraphics[width=0.55\linewidth, valign=c]{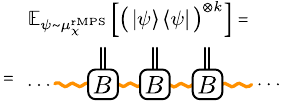}
\end{equation}
where the MPS replicated blocks are defined as~\cite{Haag_2023}
\begin{equation}\label{eq:mps_block_bulk}
\includegraphics[width=0.6\linewidth, valign=c]{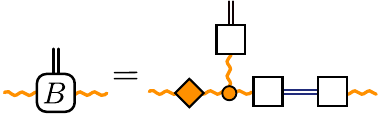}
\end{equation} 
Note that auxiliary indices are contracted within the matrices $B$, which have size $k!$, independently of $\chi$~\cite{Haag_2023}. Therefore one can easily access the large $\chi$ limit. 

By examining explicitly the block Eq.~\ref{eq:mps_block_bulk} for $k=1$, one can easily show that RMPS are $1$-design 
~\cite{Garnerone_2010_2} for any $\chi \geq 1$. However, RMPS fail to be quantum $2$-designs~\cite{Haferkamp_2021}, since for any fixed $\chi$ they have inherently low-entanglement ($\mathbb{E}_{\psi \sim \murmps} \left[ \Pur_A(\ket{\psi}) \right] \geq \chi^{-1}$) while for Haar states the average purity is exponentially small (Eq.~\ref{eq:averagepurity}).

Finally we mention that RMPS can be prepared with precision $\epsilon$ using circuits of depth $\Order(\log(N/\epsilon))$~\cite{Malz_2024}. For translationally invariant MPS, the preparation can be improved using adaptive circuits that achieve a constant depth, independent of $N$~\cite{Smith_2024}. \\

\section{RMPS with open boundary conditions}\label{sec:rmpsobc}

Here, we exemplify the open boundary conditions used in the quantum circuit representation of an MPS at finite system size $N$. In this section only, we will use the graphical convention of using single, double, triple, etc. lines to represent auxiliary bonds with dimensions $2,4,8...$. For illustrative purposes, we consider a particular case with $N=6$ qubits and maximum bond dimension $\chi=8$. A right-normalized MPS is therefore~\cite{Schollwock_2011}
\begin{equation}
\includegraphics[width=0.55\linewidth, valign=c]{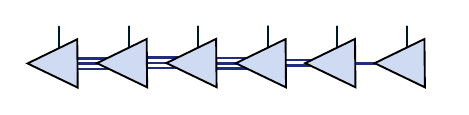}
\end{equation}
Note that it is necessary to impose a bond dimension that consecutively decreases by a factor of $2$ for each of the last $\log_2 \chi - 1$ bonds to the right. In fact, if this were not the case, the right normalization of the MPS tensors could not be fulfilled. At this point, MPS isometries $\mathbb{A}$ can be embedded into unitary gates provided an adequate number of qubits initialized to $\ket{0}$ are supplied.
This is done as follows:
\begin{equation}
\includegraphics[width=0.60\linewidth, valign=c]{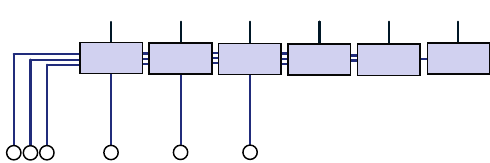}
\end{equation}
Observe that last $\log_2$ tensors do not need to be expanded, as they are already square unitary matrices. 
Finally, the circuit can be reshaped to obtain a staircase geometry: 
\begin{equation}\label{eq:circuitfinitesize}
\includegraphics[width=0.60\linewidth, valign=c]{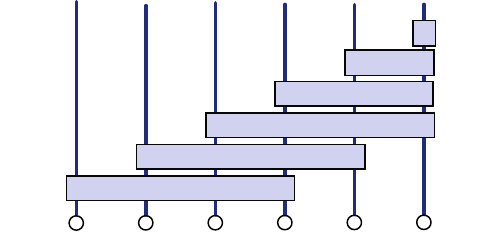}
\end{equation}
At this point, RMPS at finite system size are defined by using Haar-distributed unitary gates in the circuit described by Eq.~\ref{eq:circuitfinitesize}. Note that this construction results in normalized states, as a unitary circuit is applied to an initially normalized state $\ket{\pmb{0}}$. Additionally, observe that when $\chi = d = 2^N$, we obtain the $N$-qubit Haar distribution, because the unitary gate at the first layer covers all qubits in this case.

\section{Magic of RMPS} 

The goal of this section is to determine the average value of the linearized $n-$SRE 
$m_n(\ket{\phi}) = ||\Pi_{\phi}||_n^n$
over the ensemble of RMPS~\footnote{Similar computations have been performed in Ref.~\cite{Chen_2022,Haag_2023}}. However before of proceeding it is useful to obtain the Haar averaged value of $m_n$. This can be easily evaluated within the Weingarten calculus, as shown in Appendix~\ref{sec:haar_averages}. The result is ($n=2,3$)
\begin{align}\label{eq:haaraveragemagic}
    \begin{split}
    \Ex_{\phi\sim \muH}[m_2(\ket{\phi})] &= \frac{1}{d^2} 
    \bigg( 1 + 3 \frac{d-1}{d+3} \bigg) \\
    \Ex_{\phi\sim \muH} [m_3(\ket{\phi})] &= \frac{1}{d^3} 
    \bigg( 1 + 
    \frac{15(d-1)}{(3 + d) (5 + d)} \bigg) \, ,
    \end{split}
\end{align} 
for a system of Hilbert space dimension $d$. 

Now, we consider the simple case of 
RMPS with bond dimension $1$, i.e. random product states. We have ($\mu_1 = \mu_{\chi = 1}$):
\begin{equation}
    \Ex_{\phi\sim \mu_1}[m_n(\ket{\phi})] = \prod_{i=1}^N \bigg( 2^{-n}\sum_{\sigma_i \in \tilde{\mathcal{P}}_1} \Ex_{\phi_i \sim \muH}[ \braket{\phi_i|\sigma_i|\phi_i}^{2n} ] \bigg) \, ,
\end{equation}
where for last equality, we utilized the factorization over qubits $i = 1 ... N$ of both the product state $\ket{\phi}$ and the Pauli sum, explicitly leveraging the assumption of local dimension $2$. Thus, we find
\begin{equation}\label{eq:factorization_product}
    \Ex_{\phi\sim \mu_{1}}[m_n(\ket{\phi})] =  \big( \Ex_{\phi_i\sim \muH}[m_n(\ket{\phi_i})] \big)^N
\end{equation}
By setting $d=2$ in Eqs~\ref{eq:haaraveragemagic} and
substituting the result into Eq.~\ref{eq:factorization_product}, we obtain
\begin{equation}
    \Ex_{\phi\sim \mu_{1}}[m_2(\ket{\phi})] = \frac{1}{d^2} \bigg( \frac{8}{5} \bigg)^N \quad
    \Ex_{\phi\sim \mu_{1}}[m_3(\ket{\phi})] = \frac{1}{d^3} \bigg( \frac{10}{7} \bigg)^N \, .
\end{equation}
Notice that these values are exponentially larger in $N$ compared to the full system Haar values (Eq.~\ref{eq:haaraveragemagic}). Consequently for large $d$
\begin{align}\label{eq:randomproductstatesaveragemagic}
    \begin{split}     
    \delta^{(2)}_{\chi=1} \simeq \bigg( \frac{8}{5} \bigg)^N \qquad \delta^{(3)}_{\chi=1} \simeq \bigg( \frac{10}{7} \bigg)^N \, ,
    \end{split}
\end{align}
where we introduced the quantity
\begin{equation}\label{eq:linearizednmagicdiff}
\delta^{(n)}_{\chi} = d^n \big(\Ex_{\phi\sim \murmps}[m_n(\ket{\phi})] - \Ex_{\phi\sim \muH}[m_n(\ket{\phi})] \big)
\end{equation}
which measures the average deviation of the $n-$linearized magic from that of Haar states. The factor $d^n$ in Eq.~\ref{eq:linearizednmagicdiff} is included because, when it multiplies the Haar term, it results in a quantity of order of $\Order(1)$. Indeed (see Eq.~\ref{eq:haaraveragemagic}) for large $d$ one has: $d^2 m_2(\ket{\phi}) \simeq 4$ and $d^3 m_3(\ket{\phi}) \simeq 1$.

Finally, we consider the general case of RMPS with bond dimension $\chi$. We have
\begin{equation}
    d^n \Ex_{\mathcal{\psi} \sim \murmps}[m_n(\ket{\psi})] = \sum_{\pmb{\sigma}} \Tr[ \pmb{\sigma}^{\otimes k} \Ex_{\mathcal{\psi} \sim \murmps}[ \big( \ket{\psi} \bra{\psi} \big)^{\otimes k} ] ]\, ,
\end{equation}
where $k=2n$. Local Pauli operators $\sigma_i^{\otimes k}$ can be folded as vectors and contracted with physical indices of the block $B$ in Eq.~\ref{eq:mps_average_k_bulk}. We thus obtain
\begin{equation}\label{eq:mps_average_k_magic_bulk}
\includegraphics[width=0.73\linewidth, valign=c]{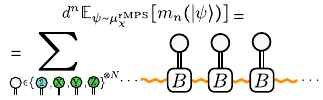}
\end{equation}
After factorizing the Pauli sum, one can introduce the following transfer matrix $\mathcal{T}$
\begin{equation}\label{eq:transfermatrix}
\includegraphics[width=0.65\linewidth, valign=c]{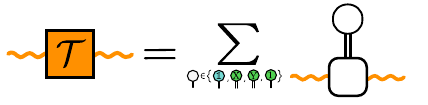} \, ,
\end{equation}
so that Eq.\ref{eq:mps_average_k_magic_bulk} becomes
\begin{equation}\label{eq:transfermatrix2}
\includegraphics[width=0.95\linewidth, valign=c]{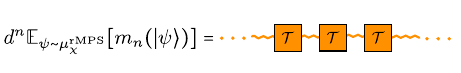} \, .
\end{equation}
Matrix elements $\mathcal{T}_{\sigma \pi}$ of the transfer matrix can be determined through symbolic computation, and consist of rational functions of $\chi$. Boundary conditions can be implemented in two distinct ways:
\begin{equation}\label{eq:pbc_or_obc}
\includegraphics[width=0.45\linewidth, valign=c]{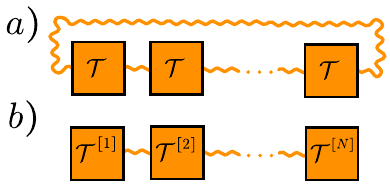} \, ,
\end{equation}
namely periodic boundary conditions (PBC), $a)$, or open boundary conditions (OBC), $b)$. We shall detail the two calculations in the upcoming subsections. The main difference between the two is that evaluating the average SRE as in Eq.~\ref{eq:pbc_or_obc} $a)$ is not exact, as for finite $N$, RMPS obtained with PBC are not properly normalized~\cite{Haferkamp_2021}. Instead introducing OBC automatically enforce normalisation (see Sec.~\ref{sec:rmpsobc}).

\subsection{Magic in RMPS with PBC}

The first approach leads to $d^n \Ex_{\phi\sim \murmps}[m_n(\ket{\phi})] = \Tr[\mathcal{T}^N]$, which is easy to handle, since the calculation boils down to finding eigenvalues $\lambda_{\sigma}$ of matrix $\mathcal{T}$ in the bulk. For convenience, we consider the eigenvalues to be sorted in descending order ($\lambda_{1} \geq \lambda_{2} \geq ... \geq \lambda_{k!}$).

\begin{figure}[ht!]
\includegraphics[width=0.65\linewidth]{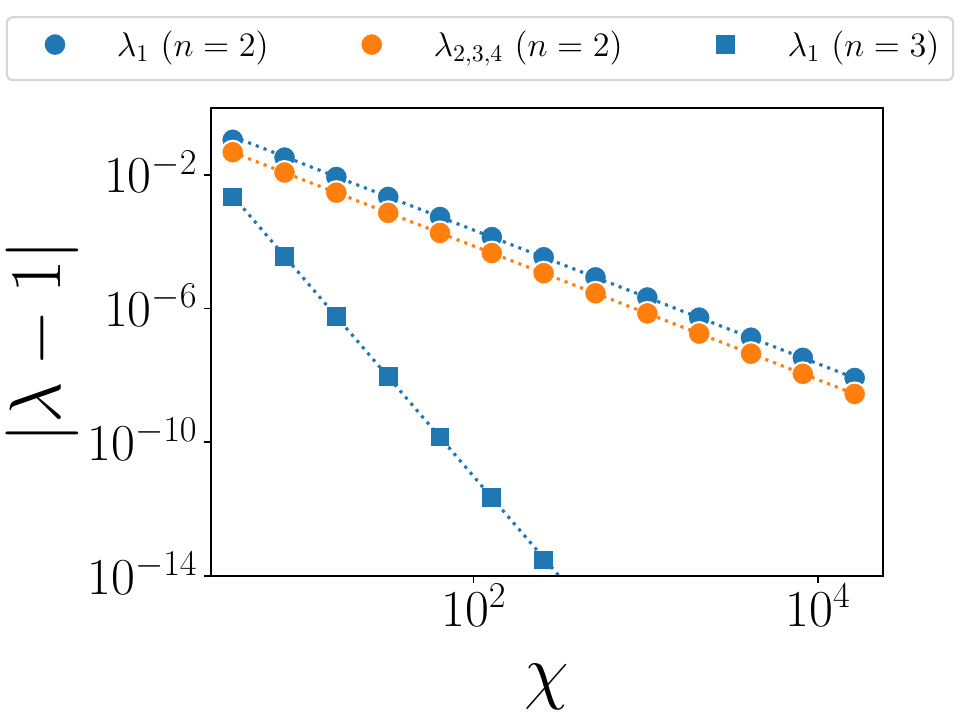}
\caption{Leading eigenvalues of the transfer matrix $\mathcal{T}$ in the bulk for $n=2$ (cirles) and $n=3$ (squares). Straight lines represent leading term for large $\chi$ extracted from analytical expansion ($n=2$) or from a linear fit ($n=3$). \label{fig:app_1}}
\end{figure}

For $n=2$ ($k=4$ replicas), the $\lambda_{\sigma}$ can be obtained using symbolic computation routines of Wolfram Mathematica. It is possible to perform an expansion for large $\chi$ and identify the leading eigenvalues. We find four leading eigenvalues converging to $1$ in the large $\chi$ limit. These are:
\begin{align}\label{eq:analyticalexpansioneigenvalues}
\begin{split}
    \lambda_1 & = 1 + \frac{9}{4 \chi^2} -\frac{171}{16 \chi^4} + \frac{5265}{64 \chi^6} + \Order\left(\frac{1}{\chi^8} \right) \\  
    \lambda_{2,3,4} & = 1-\frac{3}{4 \chi^2}-\frac{3}{16 \chi^4}-\frac{3}{64 \chi^6} + \Order\left(\frac{1}{\chi^8} \right) \\       
\end{split} 
\end{align}
All the other eigenvalues $\lambda_{\sigma}$ ($\sigma >4$) converge either to $1/2$ or to $1/4$. In Fig.~\ref{fig:app_1}, we show the deviation of $\lambda_1$ and $\lambda_{2}=\lambda_{3}=\lambda_{4}$ from $1$ as a function of $\chi$ for various values of $\chi$. Additionally, we depict the leading term for large $\chi$, which in both cases scales as $\Order(\chi^{-2})$.

For $n=3$ ($k=6$ replicas), an exact symbolic computation of the eigenvalues is not feasible, as $\mathcal{T}$ has a dimension of $720 \times 720$. However, by evaluating numerically the eigenvalues of $\mathcal{T}$ for certain fixed values of $\chi$, we observe a single leading eigenvalue $\lambda_1$ converging to $1$ from above in the large $\chi$ limit. In this case, we use a linear fit to obtain the scaling, finding 
 \begin{equation}
     \lambda_1 \simeq 1 + a \chi^{-6} \, ,
 \end{equation}
with $a \simeq 9.70$ (see Fig.~\ref{fig:app_1}). All the other eigenvalues $\lambda_{\sigma}$ ($\sigma > 1$) converge either to $1/2$, $1/4$, $1/8$ or $1/16$.

Using the results for $n=2$, we find:
\begin{equation}
     \Tr[\mathcal{T}^N] = \sum_{\sigma=1}^{k!} \lambda_{\sigma}^N \simeq \lambda_1^N + 3 \lambda_2^N \, ,
\end{equation}
a part for exponentially small corrections due to the others eigenvalues. By inserting Eq.~\ref{eq:analyticalexpansioneigenvalues} and taking only the leading terms for large $N,\chi$ and small $N/\chi^2$, we finally get 
\begin{equation}
     \Tr[\mathcal{T}^N] \simeq 4 + \frac{27}{8} \frac{N^2}{\chi ^4} \, .
\end{equation}
If we define $\delta^{(n)}_{\chi}$ as in Eq.~\ref{eq:linearizednmagicdiff}, we assume $d^2 \Ex_{\phi\sim \murmps}[m_2(\ket{\phi})] \simeq \Tr[\mathcal{T}^N]$ and we use Eq.~\ref{eq:haaraveragemagic} for the Haar value, we obtain 
\begin{equation}\label{eq:delta2_pbc}
\delta^{(2)}_{\chi} \sim \Order \left( \frac{N^2}{\chi^4} \right)  \, . 
\end{equation}
This implies that when $\chi \gg N^{1/2}$, the 2-linearized magic of the RMPS converges to that of Haar, i.e.\ $\delta^{(2)}_{\chi}$ approaches $0$. 
Similarly, one finds 
\begin{equation}\label{eq:delta3_pbc}
\delta^{(3)}_{\chi} \sim \Order \left( \frac{N}{\chi^6} \right)  \, . 
\end{equation}
i.e.\ the 3-linearized magic converges for $\chi \gg N^{1/6}$.

\begin{figure}[h!]
\hspace{0.8 cm} 
\begin{flushleft}
\hspace{6 mm} \includegraphics[width=0.90\linewidth]{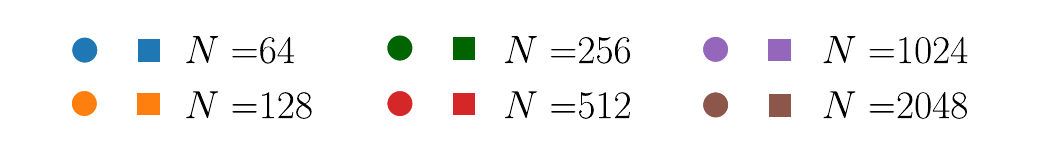} \\
\end{flushleft}
\vspace{-5 mm}
\includegraphics[width=0.52\linewidth]{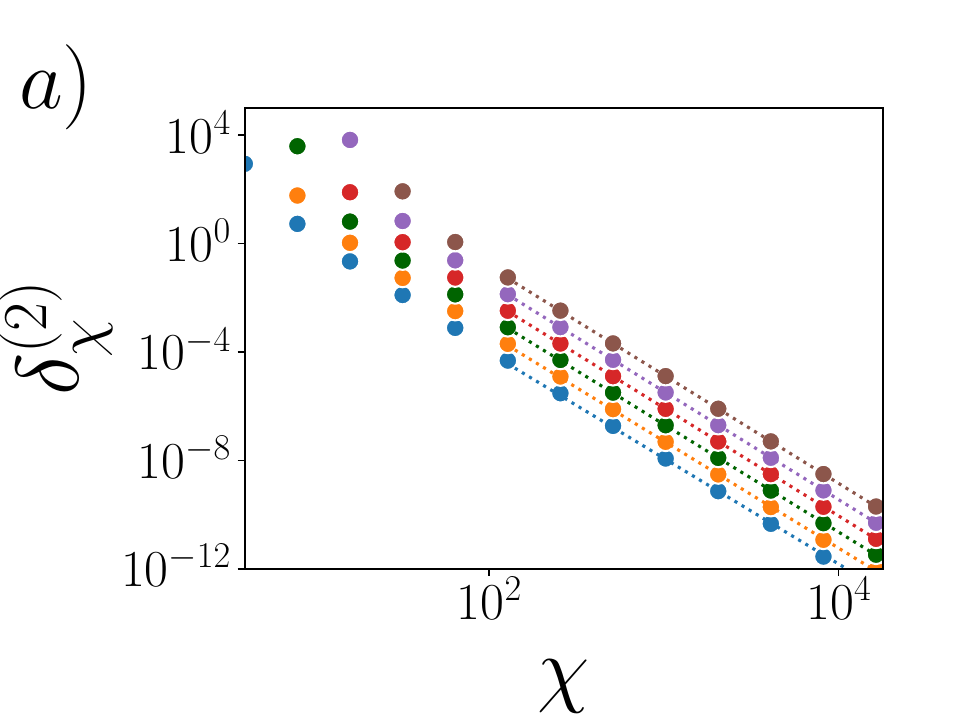}%
\includegraphics[width=0.52\linewidth]{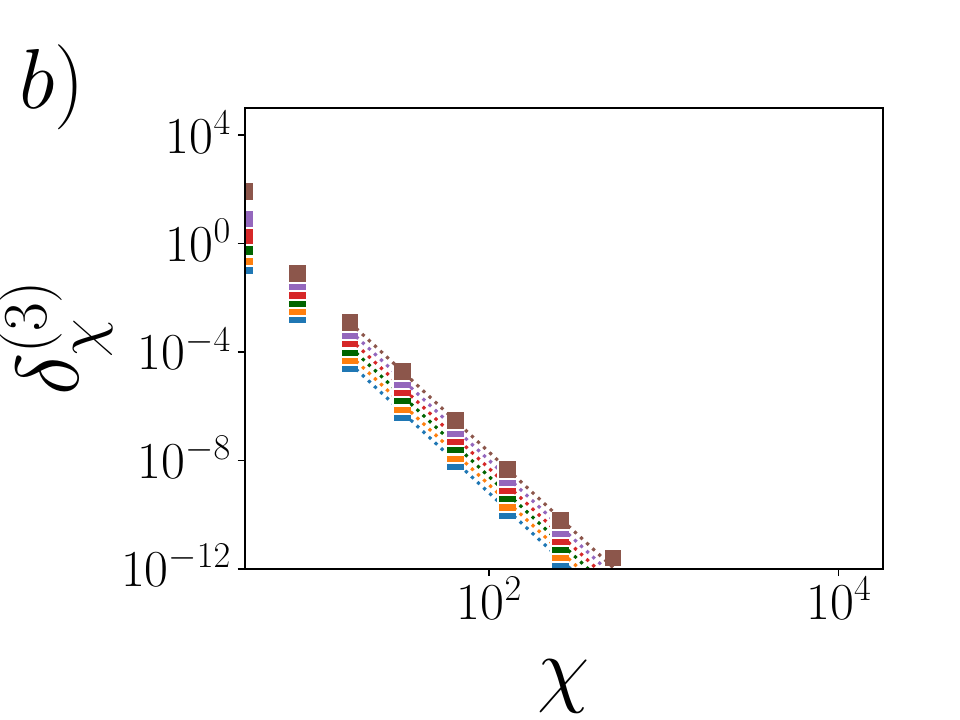} \\
\includegraphics[width=0.52\linewidth]{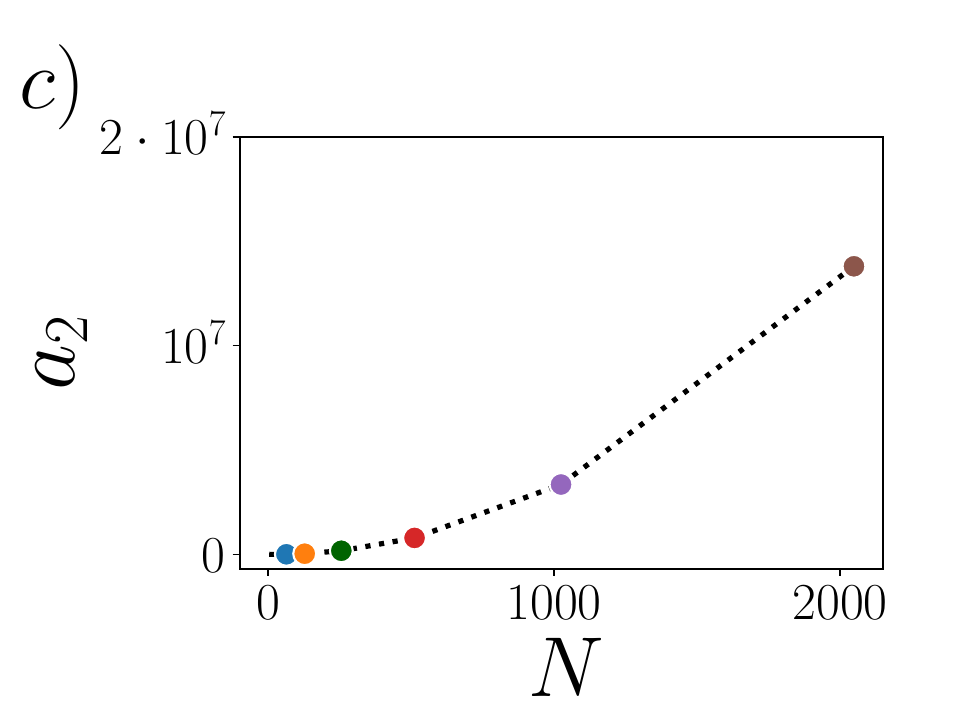}%
\includegraphics[width=0.52\linewidth]{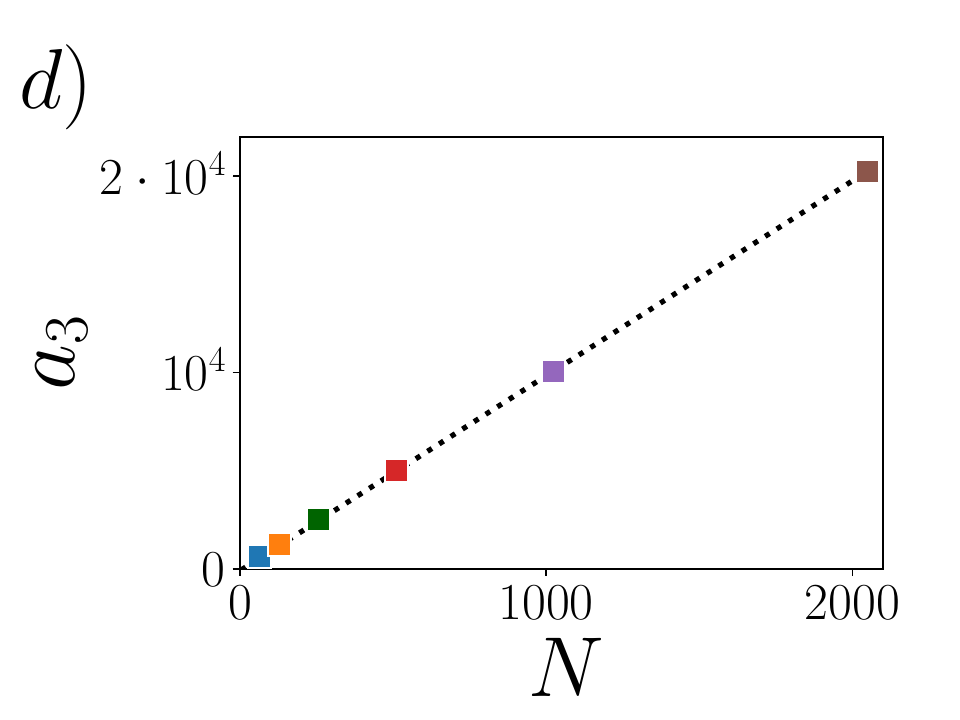} \\
\caption{Magic of RMPS for periodic boundary conditions, see Eq.~\ref{eq:pbc_or_obc}$a)$.
Upper panels: magic deviation from Haar $\delta_{\chi}^{(n)}$ for RMPS at finite size $N$ and $n=2$ ($a$), $n=3$ ($b$). Here, we employ periodic boundary conditions, see Eq.~\ref{eq:pbc_or_obc} $a)$. Dotted lines represent the linear fit in log log scale (Eq.\ref{eq:linear_fit_pbc}). Lower panels: coefficients $a_n$ as a function of $N$ for $n=2$ ($c$), $n=3$ ($d$).}
\label{fig:app_2_pbc}
\end{figure}

To confirm these findings, we also compute $\Tr[\mathcal{T}^N] = \sum_{\sigma=1}^{k!} \lambda_{\sigma}^N$
by numerically extracting the eigenvalues $\lambda_{\sigma}$. In Fig.~\ref{fig:app_2_pbc}, panels $a)$ and $b)$, display the computed values of $\delta^{(n)}_{\chi}$ for increasing values of $\chi$ and $n=2,3$. Through a linear fit (dotted lines), we find
\begin{equation}\label{eq:linear_fit_pbc}
\delta^{(2)}_{\chi} \simeq a_2 \chi^{-4} \qquad \delta^{(3)}_{\chi} \simeq a_3 \chi^{-6} \, .
\end{equation}
Subsequent fitting (see panels $c)$ and $d)$) reveals that the fitting coefficients scale as $a_2\propto N^2$ and $a_3\propto N$ respectively. Collectively, these results corroborate the analytical predictions specified in Eqs.~\ref{eq:delta2_pbc} and \ref{eq:delta3_pbc}.

\subsection{Magic in RMPS with OBC}

\begin{figure}[h!]
\hspace{0.8 cm} 
\begin{flushleft}
\hspace{6 mm} \includegraphics[width=0.90\linewidth]{legend.pdf} \\
\end{flushleft}
\vspace{-5 mm}
\includegraphics[width=0.52\linewidth]{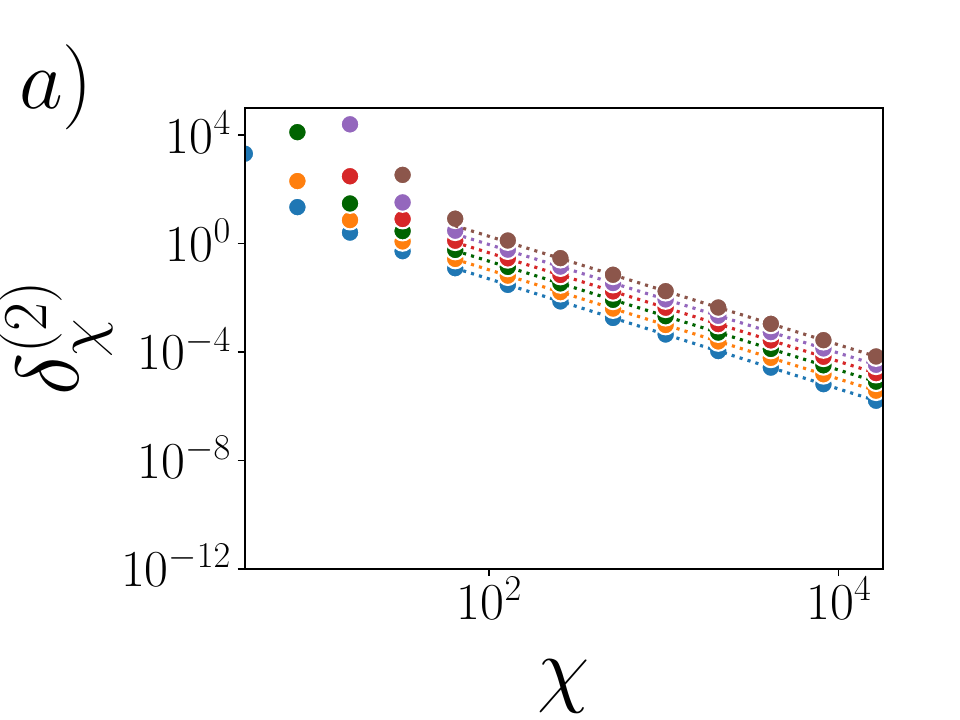}%
\includegraphics[width=0.52\linewidth]{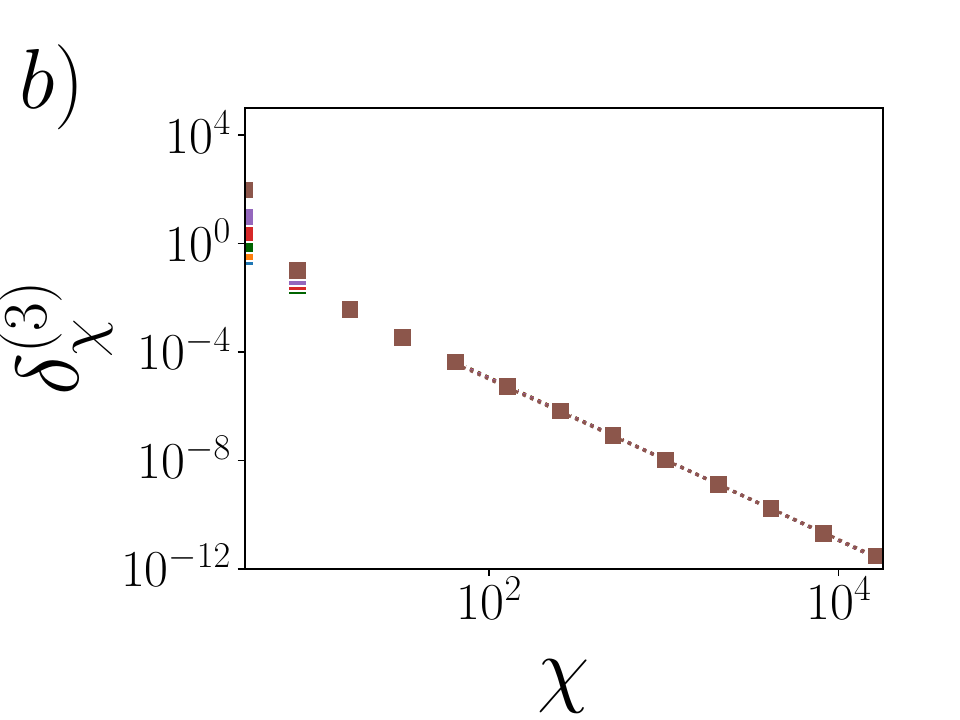} \\
\includegraphics[width=0.52\linewidth]{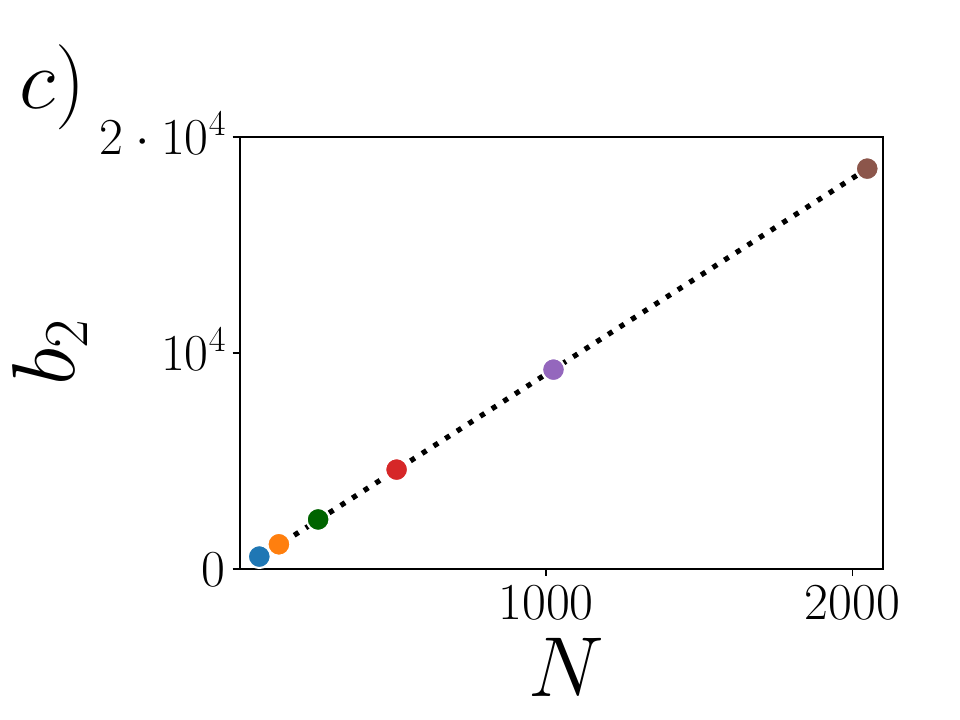}%
\includegraphics[width=0.52\linewidth]{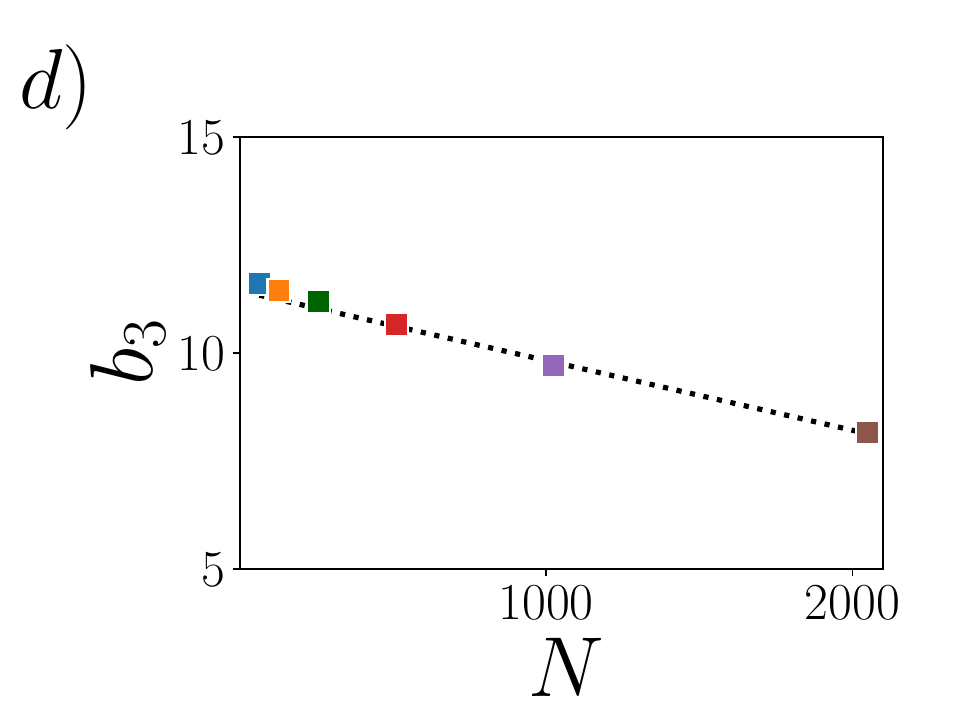} \\
\caption{Magic of RMPS with open boundary conditions, see Eq.~\ref{eq:pbc_or_obc}$b)$.
Upper panels: magic deviation from Haar $\delta_{\chi}^{(n)}$ for RMPS at finite size $N$ and $n=2$ ($a$), $n=3$ ($b$). Dotted lines represent the linear fit in log log scale (Eq.\ref{eq:linear_fit_obc}). Lower panels: coefficients $b_n$ as a function of $N$ for $n=2$ ($c$), $n=3$ ($d$).}
\label{fig:app_2_obc}
\end{figure}

In order to obtain properly normalized RMPS, we consider open boundary condition. This can be implemented as discussed in Sec.~\ref{sec:rmpsobc}. In this scenario, $\mathcal{T}$ varies across sites, and the SRE is obtained by multiplying the matrices as in Eq.\ref{eq:pbc_or_obc} $b)$. We numerically evaluate such product for RMPS with $\chi \in [2, 16384]$, $N \in [4, 2048]$. In Fig.~\ref{fig:app_2_obc}, panels $a)$ and $b)$, we plot the magic deviation from Haar $\delta_{\chi}^{n}$ ($n=2,3$). 

Through a linear fit (dotted lines), we find 
\begin{equation}\label{eq:linear_fit_obc}
\delta^{(2)}_{\chi} \simeq b_2 \chi^{-2} \qquad \delta^{(3)}_{\chi} \simeq b_3 \chi^{-3} \, .
\end{equation}
Subsequent fitting (see panels $c)$ and $d)$) reveals that $b_n\propto N$ for both $n=2$ and $n=3$. This reveals that the correct scaling of the magic deviation from Haar {for open boundary conditions} is
\begin{equation}
\delta^{(2)}_{\chi} \sim \Order \left( \frac{N}{\chi^2} \right)  \quad 
\delta^{(3)}_{\chi} \sim \Order \left( \frac{N}{\chi^3} \right)  \, . 
\end{equation}
To alternatively confirm this scenario, we simulate the evolution of a circuit with 2-qubit Haar random unitaries, allowing the MPS bond dimension to grow up to $\chi$ (see Fig.~\ref{fig:1}$a$). Even though, strictly speaking, the resulting ensemble of states does not exactly correspond to the usual definition of RMPS, measuring $\delta^{(n)}_{\chi}$ ($n=2,3$) reveals power law decays of $\chi^{-2}$ and $\chi^{-3}$ consistent with predictions from finite-size RMPS calculations (see Fig.~\ref{fig:1}~$b$). \\

\section{Clifford enhanced RMPS}--
We are now in position to introduce a novel ensemble of quantum states, dubbed $\Cliff$MPS. These are defined as 
\begin{equation*}
        \Cliff \text{MPS} = \{ \ket{\psi} = \mathcal{U}_c \ket{\phi}_{\chi} , \, \mathcal{U}_c \in \Cliff(d) \, \text{ and } \ket{\phi}_{\chi} \in \text{MPS} \} \, ,
\end{equation*}
i.e.\ by applying a Clifford circuit to an MPS with maximum bond dimension $\chi$. Note that arbitrary $\Cliff$MPS can be prepared using circuits of depth $\Order(N)$, as this is the depth required by generic Clifford unitaries~\cite{maslov2018shorter}. The depth of $\Cliff$MPS can be reduced to $\Order(\log(n))$ by restricting to Clifford constructions that form $2$-designs~\cite{Cleve_2015} and normal MPS~\cite{Malz_2024,Smith_2024}. 

Hereafter, we use $\Ex[...]$ to denote average over $\Cliff$MPS, i.e.\ over both Clifford $\mathcal{U}_c$ and MPS $\ket{\phi}_{\chi}$, the latter equipped with the measure $\murmps$. 

First, it is trivial to show that $\Cliff$MPS inherit from $\Cliff(d)$ the property of being a quantum $3$-design. Indeed, for any $k \leq 3$, we have
\begin{align}
    \begin{split}
            &\Ex[(\ket{\psi} \bra{\psi})^{\otimes k}] = \Ex_{\phi \sim \murmps} \big[ \Ex_{\mathcal{U} \sim \muCl}[(\mathcal{U}_c^{\dag} \ket{\phi} \bra{\phi} \mathcal{U}_c)^{\otimes k}] \big] = \\ 
            &=\Ex_{\phi \sim \murmps} \big[ \Ex_{\mathcal{U} \sim \muH}[(\mathcal{U}^{\dag} \ket{\phi} \bra{\phi} \mathcal{U})^{\otimes k}] \big] = \Ex_{\psi \sim \muH}[(\ket{\psi} \bra{\psi})^{\otimes k}] \, .
    \end{split}
\end{align}
Thus, to detect deviations of the $\Cliff$MPS ensemble from Haar, we must study fourth moments. In order to quantify these deviations, we evaluate the $\Cliff$MPS $4$-frame potential
\begin{align}
    \begin{split}
    \mathcal{F}^{(4)}_{\Cliff \text{MPS}}=\Ex \left[\Tr[(\mathcal{U}_c^{\dag} \ket{\phi} \bra{\phi} \mathcal{U}_c)^{\otimes 4} (\mathcal{V}_c^{\dag} \ket{\varphi} \bra{\varphi} \mathcal{V}_c)^{\otimes 4}]\right] \, ,
    \end{split}
\end{align}
where we average over Clifford $\mathcal{U}_c,\mathcal{V}_c$ and RMPS $\ket{\phi},\ket{\varphi}$. By using the normalized Schatten-$2$ distance from Haar~\cite{ippoliti2023dynamical,Ippoliti_2022} $\Delta^{(k)}=(\mathcal{F}^{(k)} / \mathcal{F}^{(k)}_{\text{H}} - 1)^{1/2}$ (which is an upper bound to the trace-distance deviation), and using Eq.~\ref{eq:cliff4moment}, we get 
\begin{align}\label{eq:framepotentialdistcmps}
    \begin{split}
    \Delta^{(4)} = \frac{d+3}{d} \frac{1}{\big(4(d-1)(4 + d)\big)^{1/2}} \delta^{(2)}_{\chi} \sim \frac{1}{2d} \delta^{(2)}_{\chi} \, .
    \end{split}
\end{align}
Another way to quantify the closeness of $\Cliff$MPS to Haar is calculating the fluctuations of purity (Eq.~\ref{eq:purfluc}) and verifying how these interpolate between the Clifford regime 
$\Delta^2 \Pur_A|_{\text{STAB}} \sim \Order(d^{-1})$
and the Haar regime $\Delta^2 \Pur_A|_{\text{H}} \sim \Order(d^{-2})$ (Eqs.~\ref{eq:purfluccliff},~\ref{eq:purfluchaar}). By exploiting again Eq.~\ref{eq:cliff4moment}, one gets (see Appendix~\ref{sec:clifford_averages} for details)
\begin{equation}\label{eq:purityfluctfinal}
\Delta^2 \Pur_A(\ket{\psi}) = \Delta^2 \Pur_A(\ket{\psi})\big\rvert_{\text{H}} + \frac{(d-1)}{d(d+1)(d+2)} \delta^{(2)}_{\chi} \, .
\end{equation}
In summary, both the computation of the $4$-frame potential and the analysis of purity fluctuations boil down to evaluating $\delta_{\chi}^{(2)}$. Consequently, the way the ensemble $\mathcal{U}_c \ket{\phi}$ becomes close to Haar random states depends on the magic of the states $\ket{\phi}$ (in our case RMPS).
For $\chi=1$, one finds  $\delta_{\chi}^{(2)} \sim \Order(d^{\log_2 (8/5)})$ (see Eq.~\ref{eq:randomproductstatesaveragemagic}) and therefore, $\Delta^2 \Pur_A(\ket{\psi}) \sim \Order(d^{1 - \log_2 5})$. This indicates that random product states enhanced by Clifford circuits still exhibit exponentially larger purity fluctuations compared to those of Haar random states. However, as shown in previous section, for any fixed $N$, $\delta^{(2)}_{\chi}$ can be made arbitrarily small by increasing $\chi$. The closeness to a $4$-design $\Delta^{(4)}$ can be reduced in the same way. More specifically, for a given $N$ and any $\epsilon > 0$, achieving $\Delta^{(4)} < \epsilon$ requires that $\chi > \frac{1}{\sqrt{2 \epsilon}} \sqrt{\frac{N}{d}}$. \\

\section{Conclusions and Outlook} 
We have given new insights into the relationship between two fundamental quantum resources: entanglement and magic. Specifically, we demonstrated that MPS, which inherently possess bounded entanglement, when enhanced by the application of Clifford (magic-free) unitaries, can yield highly non-trivial quantum states that exhibit characteristics akin to those of generic Haar states. We denote by $\Cliff$MPS the ensemble of states $\ket{\psi} = \mathcal{U}_c \ket{\phi}_{\chi}$, where $\ket{\phi}_{\chi}$ is an MPS and $\mathcal{U}_c$ a Clifford. By appropriately adjusting the MPS bond dimension $\chi$, one can achieve an approximate $4$-design with arbitrary precision, which scales as $\chi^{-2}$. Although we do not provide arguments in this regard, we believe that this result can be extended to the case of $6$-design, obtaining a similar power-law scaling for $\Delta^{(6)}$~\cite{Leone_2024_mail}.

The question of whether $\Cliff$MPS representation could enhance classical simulations of many-body quantum systems or quantum circuits by effectively combining the MPS and Clifford (tableau) formalism remains open~\cite{Paviglianiti_2024}. 
While sampling $\Cliff$MPS in the computational basis is likely challenging~\cite{Yoganathan_2019}, one can efficiently compute the expectation value of any Pauli string $\pmb{\sigma}$. This efficiency arises as $\braket{\psi|\pmb{\sigma}|\psi} = \braket{\phi|\pmb{\sigma}'|\phi}$, where $\sigma' = \mathcal{U}_c^\dagger \pmb{\sigma} \mathcal{U}_c$. We can efficiently compute $\braket{\psi|\pmb{\sigma}|\psi}$ by using the Clifford tableau formalism to transform $\pmb{\sigma}$ into $\pmb{\sigma}'$. Subsequently, we contract the MPS to evaluate $\braket{\phi|\pmb{\sigma}'|\phi}$, which can be achieved in $\Order(N \chi^3)$ time~\cite{Schollwock_2011}. In Appendix~\ref{sec:cooling}, we exemplify how a circuit of two-qubit Clifford unitaries can be used to largely decrease the entanglement of a highly entangled state. The Clifford transformation is obtained by applying the entanglement cooling algorithm, initially studied in Ref.~\cite{True_2022}, which optimizes local two-qubit Clifford gates to minimize the entanglement entropy of the state. The target state $\ket{\psi}$ we consider is obtained by evolving an initial product state with a random $T$-doped Clifford circuit. Entanglement cooling shows how  elements in the ensemble $\mathcal{U}_c \ket{\phi}$ can be successfully used to reproduce a complex states $\ket{\psi}$. Extending the entanglement cooling to target MPS with bond dimension $\chi$ polynomial in $N$, as well as developing analogous MPS-Clifford techniques, represents a formidable challenge for future research. Such approaches, if realized, could prove particularly effective for tasks that require only the computation of Pauli expectation values, such as determining ground states or simulating the time evolution of Hamiltonians~\cite{cao2019quantum,Daley_2022,bauer2023quantum}. Finally, it would be interesting to understand whether the $\Cliff$MPS representation of a given state $\ket{\psi}$ can be efficiently learned~\cite{Huang_2024}.\\

\paragraph{Acknowledgments.}

We are particularly grateful to L. Piroli for useful advices on RMPS calculations. We also thank L. Leone and S. Oliviero for inspiring discussions. 
We also want to thank S. Gopalakrishnan,   A. Paviglianiti, M. Collura, A. Hamma, L. Tagliacozzo and A. De Luca for inspiring discussions and for collaborations on topics connected with this work. This work was supported by ANR-22-CPJ1-0021-01 and ERC Starting Grant 101042293
(HEPIQ).\\

\textit{Note added---}  While finalizing this work, we became
aware of a related study on the convergence of magic with bond dimension in MPS representing ground states. The paper, authored by M.~Frau, P.S.~Tarabunga, M.~Collura, M.~Dalmonte, E.~Tirrito, appeared on the same arxiv posting~\cite{frau2024non}. \\

\bibliographystyle{apsrev4-2}
\bibliography{bib}

\appendix
\appendixtitleon
\appendixtitletocon

\setcounter{secnumdepth}{3}
\begin{appendices}

\section{$\Cliff$MPS and entanglement cooling}\label{sec:cooling}

We now describe and test an algorithm aimed at finding the optimal Clifford unitary $\mathcal{U}_c^{\dag}$ to disentangle a given state $\ket{\psi}$. The goal is to find an approximate $\Cliff$MPS decomposition, i.e.\
\begin{equation}
   \ket{\psi} \simeq \mathcal{U}_c \ket{\phi}_{\chi} \,
\end{equation}
where $\ket{\phi}_{\chi}$ is a suitable MPS, with (small) bond dimension $\chi$. The approach is known as entanglement cooling~\cite{True_2022}. Essentially, one applies two-local Clifford gates to $\ket{\psi}$, optimizing each gate to minimize the entanglement entropy between qubits $[1,i]$ and $[i+1,N]$. Given that the size of the two-qubit Clifford group is $|\Cliff(4)| = 11520$, the optimal Clifford gate can be determined numerically through a brute-force search. After identifying the optimal gate, the process is repeated sequentially for the next bonds, and additional layers of Clifford gates are optimized in a similar manner. The algorithm ultimately outputs an optimized disentangling Clifford circuit of a specific depth. Extensions of the algorithm with finite-temperature stochastic optimization of local Clifford gates can be also considered~\cite{True_2022}. \\

Here, we bench-mark the algorithm on states $\ket{\psi}$ resulting from random Clifford circuits doped with $T$ gates. Specifically, we construct $\ket{\psi}$ by alternating brick wall layers of random two-qubit Clifford gates with a single $T$ gate placed at a random position. Afterwards we apply entanglement cooling on $\ket{\psi}$, as illustrated here:
\begin{equation}\label{eq:scheme_cooling}
\includegraphics[width=0.60\linewidth, valign=c]{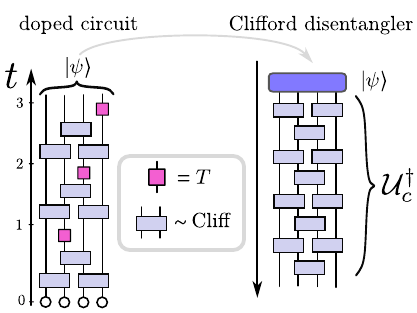} \, .
\end{equation}
Note that the discrete time of the circuit is conventionally taken to be $1$ when $v$ brick wall layers of local Clifford gates have been applied, followed by one $T$ gate. Physically, $v$ represents the circuit's light velocity ($v=1$ in the picture).

\begin{figure}[h!]
\includegraphics[width=0.9\linewidth]{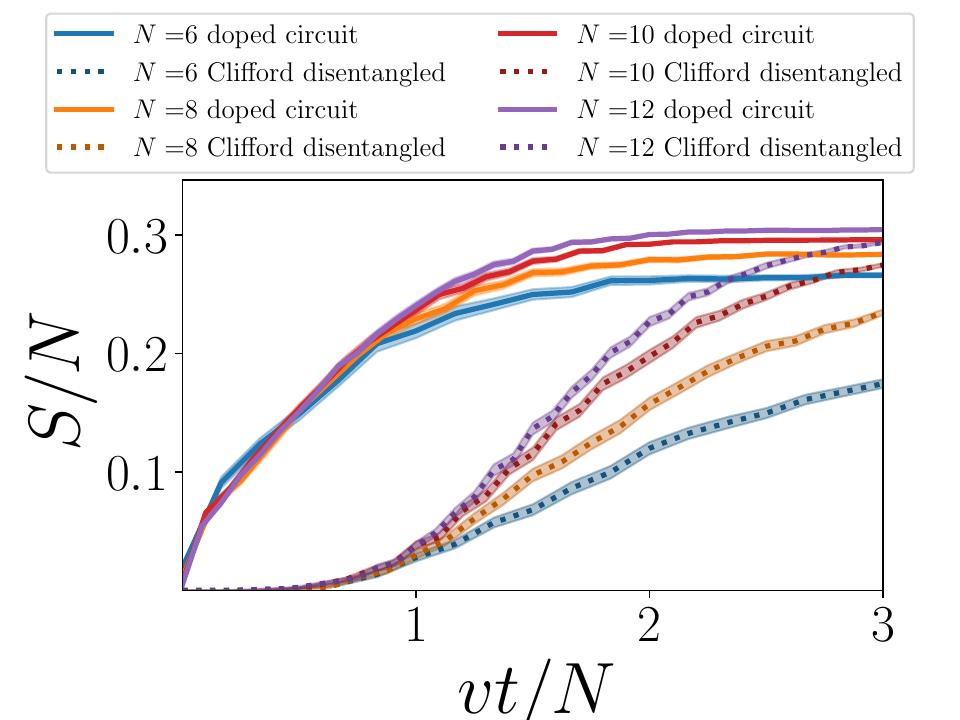} 
\caption{Entanglement cooling is applied to doped circuit (see Eq.~\ref{eq:scheme_cooling}). Continuous lines represent entanglement of the final state $\ket{\psi}$, dotted lines the entanglement of the Clifford disentangled state $\mathcal{U}_c^{\dag}\ket{\psi}$.   \label{fig:entanglement_cooling}}
\end{figure}

In Fig.~\ref{fig:entanglement_cooling}, we show results for system of size $N=6,8,10,12$ and $10^{2}$ random realizations of the doped circuit. The depth of the Clifford disentangling circuit is set to $N$. We plot the entanglement entropy of the state $\ket{\psi}$ resulting from the doped circuit (solid lines), and entanglement entropy of the disentangled state $\mathcal{U}_c^{\dag}\ket{\psi}$ (dotted lines). The entanglement entropy $S$ is maximized across system bipartitions and averaged over circuit realizations. In the plot, we scale $S$ as $S/N$, and the circuit time $t$ as $vt/N$.
Our results suggest that doped circuits with $vt/N \lesssim 1$ can be effectively disentangled using Clifford operations. However, for $vt/N \gtrsim 1$, this method becomes progressively less effective. The entanglement entropy of the disentangled state $\mathcal{U}_c^{\dag}\ket{\psi}$ appears to follow a volume law, as evidenced by the converging curves of $S/N$. This suggests that entanglement cooling may not be suitable for finding an optimal $\Cliff$MPS representation. Indeed, the findings of this paper indicate that a $\Cliff$MPS representation for generic states $\ket{\psi}$ can potentially be achieved with an MPS of bond dimension $\mathcal{O}(N^{1/2})$, corresponding to an entanglement $S \sim \mathcal{O}(\log N)$.

\section{Haar averages}\label{sec:haar_averages}

In this Appendix, we present additional calculations for the Haar average of the linearized magic and for the purity fluctuations.

\subsection{Haar average of magic}
Here we derive the average value of the linearized magic $m_n(\ket{\phi}) = ||\Pi_{\phi}||_n^n$ for Haar states. Using the Haar average formula $\Ex_{\phi \sim \muH}[(\ket{\phi}\bra{\phi})^{\otimes k}] = P^{(k)}_{\text{symm}}/ \Tr[P^{(k)}_{\text{symm}}]$, we have:
\begin{align}
    \begin{split}
    \Ex_{\phi\sim \muH}[||\Pi_{\phi}||_n^n] &= d^{-n} \sum_{\pmb{\sigma} \in \tilde{\mathcal{P}}_N} \Ex_{\phi\sim \muH}[ \braket{\phi|\pmb{\sigma}|\phi}^{2n} ] \\& = d^{-n} (\Tr[P^{(k)}_{\text{symm}}])^{-1} \sum_{\pmb{\sigma} \in \tilde{\mathcal{P}}_N} \Tr[\pmb{\sigma} P^{(k)}_{\text{symm}}] \, , \\
    \end{split}
\end{align}
where $k=2n$. The trace of an operator times the symmetric projector $P^{(k)}_{\text{symm}}$ can be systematically computed by counting the length of cycles associated to each permutation in $S_k$~\cite{Kostenberger_2021}. The result is 
\begin{align}
\begin{split}
\Tr[\pmb{\sigma} P^{(4)}_{\text{symm}}] & = \frac{1}{4!} \big( \Tr[\pmb{\sigma}]^4 + 6 \, \Tr[\pmb{\sigma}]^2 \, \Tr[\pmb{\sigma}^2] + \\ & + 3 \, \Tr[\pmb{\sigma}^2]^2 + 8 \, \Tr[\pmb{\sigma}] \, \Tr[\pmb{\sigma}^3] + 6 \, \Tr[\pmb{\sigma}^4] \big) \\
\end{split}
\end{align}
for $n=2$, and a similar (but longer) expression for $n=3$. The summation over $\pmb{\sigma} \in \tilde{\mathcal{P}}_N$ splits into the identity $\pmb{\sigma} = \Id$ plus all others $(d^2-1)$ Pauli strings $\pmb{\sigma} \neq \Id$ (which give the same contribution). Finally one obtain:
\begin{align}\label{eq:haaraveragemagic2}
    \begin{split}
    \Ex_{\phi\sim \muH}[||\Pi_{\phi}||_2^2] &= \frac{1}{d^2} 
    \bigg( 1 + 3 \frac{d-1}{d+3} \bigg) \, . \\
    \end{split}
\end{align}
For the case $n=3$ instead
\begin{align}\label{eq:haaraveragemagic3}
    \begin{split}
    \Ex_{\phi\sim \muH}[||\Pi_{\phi}||_3^3] &= \frac{1}{d^3} 
    \bigg( 1 + 
    \frac{15(d-1)}{(3 + d) (5 + d)} \bigg) \, .\\
    \end{split}
\end{align} 
Notice that in the limit of large system size one finds
\begin{equation}\label{eq:haarvaluemagic}
    \lim_{d \rightarrow \infty} \big( d^n \Ex_{\phi\sim \muH}[||\Pi_{\phi}||_n^n] \big) = \begin{cases}
        4 \text{,  if $n=2$} \\
        1 \text{,  if $n=3$} \\
    \end{cases} \, .
\end{equation}

\subsection{Haar average of purity fluctuations}\label{sec:haar_average_purity_fluc}
Here we derive the average value of purity fluctuations for Haar states, which is defined as 
\begin{equation}
    \Delta^2 \Pur_A(\ket{\psi}) \big\rvert_{\text{H}} = \mathbb{E}_{\psi\sim \muH} \left[ \Pur_A(\ket{\psi})^2 \right] - \mathbb{E}_{\psi\sim \muH} \left[ \Pur_A(\ket{\psi}) \right]^2 \, .
\end{equation}
The second term is the square of the average purity, and has already been computed in the main text  (Eq.~\ref{eq:averagepurity}), obtaining:
\begin{equation}
    \mathbb{E}_{\psi\sim \muH} \left[ \Pur_A(\ket{\psi}) \right]^2 = \bigg( \frac{d_A + d_B}{d_A d_B + 1} \bigg)^2 = \frac{4d}{(d + 1)^2} \, ,
\end{equation}
where we set $d_A=d_B=d^{1/2}$. The first term gives instead
\begin{align}
    \begin{split}
    \mathbb{E}_{\psi\sim \muH} \left[ \Pur_A(\ket{\psi})^2 \right] & = \mathbb{E}_{\psi\sim \muH} \left[ \Tr[ \rho^{\otimes 4} T^{(A)}_{2143}] \right] \\& = \frac{1}{\Tr[P^{(4)}_{\text{symm}}]} \Tr[ P^{(4)}_{\text{symm}} T^{(A)}_{2143}] = \\ &= \frac{\sum_{\pi \in S_4} \bigg( \Tr_A[ T_{\pi}^{(A)} T^{(A)}_{2143}] \Tr_B[ T_{\pi}^{(B)}] \bigg)}{d(d+1)(d+2)(d+3)} \, ,    
    \end{split}
\end{align}
where $\rho = \ket{\psi} \bra{\psi}$ and we used the fact that every permutation $T_{\pi}$ can be split as $T_{\pi} = T_{\pi}^{(A)} \otimes T_{\pi}^{(B)}$. The term $\Tr_A[ T_{\pi}^{(A)} T^{(A)}_{2143}]$ can be rewritten as $\langle \langle T_{\pi}^{(A)} |  T^{(A)}_{2143} \rangle \rangle$, where permutation operators reshaped into vector states. Consequently, one needs to calculate the overlaps of these permutation vectors. This calculation can be efficiently performed using permutation formalism, which involves merely counting the permutation cycles (see Ref.~\cite{mele_2024,Chen_2022, Haag_2023} for details). By explicitly evaluating all contributions in the sum, we obtain (again with $d_A=d_B=d^{1/2}$)
\begin{equation}
    \mathbb{E}_{\psi\sim \muH} \left[ \Pur_A(\ket{\psi})^2 \right] = \frac{2 + 18 d + 4 d^2}{(d + 1) (d + 2) (d + 3)} \, ,
\end{equation}
and finally 
\begin{equation}
    \Delta^2 \Pur_A(\ket{\psi}) \big\rvert_{\text{H}} = \frac{2(d-1)^2}{(d+1)^2(d+2)(d+3)} \, .
\end{equation}
Notice that for large $d$, $\Delta^2 \Pur_A(\ket{\psi}) \big\rvert_{\text{H}} \sim \Order(d^{-2})$.

\section{Clifford averages}\label{sec:clifford_averages}

\subsection{Frame potential}
Here we derive the expression for the frame potential $\mathcal{F}^{(4)}$ of any ensemble of states of the form $\mathcal{U}_c \ket{\phi}$. We have to evaluate 
\begin{align}
    \begin{split}    
    \mathcal{F}^{(4)}=\Ex_{\mathcal{U}_c, \mathcal{V}_c \sim \muCl} \left[\Tr[(\mathcal{U}_c^{\dag} \ket{\phi} \bra{\phi} \mathcal{U}_c)^{\otimes 4} (\mathcal{V}_c^{\dag} \ket{\varphi} \bra{\varphi} \mathcal{V}_c)^{\otimes 4}]\right] \, ,
    \end{split}
\end{align}
where $\ket{\phi}, \ket{\varphi}$ are fixed states. Now we use the expression of the $4$-fold Clifford channel
\begin{equation}
\mathbb{E}_{\mathcal{U}_c \sim \muCl} \left[ \big(\mathcal{U}_c^{\dag} \ket{\phi}\bra{\phi} \mathcal{U}_c \big)^{\otimes 4}
\right] = \alpha_{\phi} Q P^{(4)}_{\text{symm}} + \beta_{\phi} P^{(4)}_{\text{symm}}  \,
\end{equation}
\begin{equation}
    \mathbb{E}_{\mathcal{U}_c \sim \muCl} \left[ \big(\mathcal{U}_c^{\dag} \ket{\varphi}\bra{\varphi} \mathcal{U}_c \big)^{\otimes 4}
\right] = \alpha_{\varphi} Q P^{(4)}_{\text{symm}} + \beta_{\varphi} P^{(4)}_{\text{symm}}\,
\end{equation}
with the coefficients $\alpha_{\phi}, \beta_{\phi}, \alpha_{\varphi}, \beta_{\varphi}$ defined as in Eq.~\ref{eq:cliff4momentcoeff}. We now need to compute traces of products of operators like $\Tr[P^{(4)}_{\text{symm}} P^{(4)}_{\text{symm}}]$, $\Tr[P^{(4)}_{\text{symm}} Q P^{(4)}_{\text{symm}}]$, etc. The calculation simplifies greatly using the cyclicity of the trace, the fact that $P$ and $Q$ are projectors ($(P^{(4)}_{\text{symm}})^2 = P^{(4)}_{\text{symm}}$, $Q^2 = Q$), and $[Q, T_{\pi}]=0$~\cite{Leone_2021}. The remaining traces ($\Tr[P^{(4)}_{\text{symm}}]$, $\Tr[P^{(4)}_{\text{symm}} Q]$) can be evaluated explicitly using permutation formalism~\cite{Leone_2021}. The final result by normalizing with the fourth frame potential of Haar $\mathcal{F}^{(4)}_{\text{H}}$ 
\begin{align}
    \begin{split}
    \big( \Delta^{(4)} \big)^2 = \frac{\mathcal{F}^{(4)}}{\mathcal{F}^{(4)}_{\text{H}}} - 1 \\&  = \frac{\big( d^2 ||\Pi_{\phi}||_2^2 \frac{d+3}{d} - 4 \big)\big( d^2 ||\Pi_{\varphi}||_2^2 \frac{d+3}{d} - 4 \big)}{4(-1 + d)(4 + d)}  \, .
    \end{split}
\end{align}
As a particular case (including the average over the RMPS), we obtain Eq.~\ref{eq:framepotentialdistcmps}, i.e.\
\begin{align}
    \begin{split}
    \big( \Delta^{(4)} \big)^2 = \left(\frac{d+3}{d}\right)^2 \frac{1}{4(-1 + d)(4 + d)} \big( \delta^{(2)}_{\chi} \big)^2 \, .
    \end{split}
\end{align}
Instead if we set $\ket{\phi} = \ket{\varphi} = \ket{\pmb{0}}$ we obtain the frame potential of the ensemble of pure stabilizer states~\cite{Kueng_2015}. 
In this case $||\Pi_{\phi}||_2^2 = ||\Pi_{\varphi}||_2^2 = d^{-1}$, and therefore~\cite{Kueng_2015}
\begin{equation}
\mathcal{F}^{(4)}_{\text{STAB}} = \frac{30}{d (d+1) (d+2) (d+4)} \, , \qquad \big( \Delta^{(4)} \big)^2 = \frac{4(d - 1)}{(4 + d)} \, .
\end{equation}
Notice that $\big( \Delta^{(4)} \big)^2 > 0$ for any $d>1$, meaning that stabilizer states does not constitute an exact $4$-design.

\subsection{Purity fluctuations}

Here we derive the average value of purity fluctuations for any ensemble of states of the form $\mathcal{U}_c \ket{\phi}$. We proceed in close similarity to what was done in  Sec.~\ref{sec:haar_average_purity_fluc} for Haar. We have to compute $\mathbb{E}_{\mathcal{U}_c \sim \muCl} \left[ \Pur_A(\ket{\psi})^2 \right]$. By using Eqs.\ref{eq:cliff4moment},\ref{eq:cliff4momentcoeff} for the $4$-fold Clifford channel, we get
\begin{align}
    \begin{split}
    & \mathbb{E}_{\mathcal{U}_c \sim \muCl} \left[ \Pur_A(\ket{\psi})^2 \right]  = \mathbb{E}_{\mathcal{U}_c \sim \muCl} \left[ \Tr[ \rho^{\otimes 4} T^{(A)}_{2143}] \right] =  \\&  \frac{\left( \alpha_{\phi} \Tr[ P^{(4)}_{\text{symm}} T^{(A)}_{2143}] + \beta_{\phi} \Tr[Q P^{(4)}_{\text{symm}} T^{(A)}_{2143}] \right)}{\Tr[P^{(4)}_{\text{symm}}]}  \\ &= \frac{1}{d(d+1)(d+2)(d+3)} \\& \times \Big[ \sum_{\pi \in S_4} \bigg( \Tr_A[ T_{\pi}^{(A)} T^{(A)}_{2143}] \Tr_B[ T_{\pi}^{(B)}] \bigg) \\& + d^{-2} \sum_{\pmb{\sigma}} \sum_{\pi \in S_4} \bigg( \Tr_A[\pmb{\sigma}^{(A)} T_{\pi}^{(A)} T^{(A)}_{2143}] \Tr_B[\pmb{\sigma}^{(B)} T_{\pi}^{(B)}] \bigg) \Big] \, ,\\
    \end{split}
\end{align}
where $\rho = \ket{\psi} \bra{\psi} = \mathcal{U}_c^{\dag} \ket{\phi} \bra{\phi} \mathcal{U}_c$. In the second line, we split the permutation $T_{\pi}$ as $T_{\pi} = T_{\pi}^{(A)} \otimes T_{\pi}^{(B)}$ and the Pauli string as $\pmb{\sigma} = \pmb{\sigma}^{(A)} \otimes \pmb{\sigma}^{(B)}$. By explicitly evaluating the contributions given by all permutations in the sum, we obtain (again with $d_A=d_B=d^{1/2}$)
\begin{equation}
    \Delta^2 \Pur_A(\ket{\psi}) = \frac{(d-1) \left(d^2 ||\Pi_{\phi}||_2^2 +d ||\Pi_{\phi}||_2^2-2\right)}{(d+1)^2 (d+2)} \, .
\end{equation}
As a particular case we obtain the average purity of $\Cliff$MPS. 
By using the definition of $\delta_{\chi}^{(2)}$ (see Eq.~\ref{eq:linearizednmagicdiff}) we obtain  
\begin{equation}
    ||\Pi_{\phi}||_2^2 = \Ex_{\phi\sim \muH}[m_n(\ket{\phi})] + d^{-2} \delta_{\chi}^{(2)} \, ,
\end{equation}
and by replacing the Haar averaged magic with its value (Eq.~\ref{eq:haaraveragemagic2}) we finally obtain
\begin{equation}
\Delta^2 \Pur_A(\ket{\psi}) = \Delta^2 \Pur_A(\ket{\psi})\big\rvert_{\text{H}} + \frac{(d-1)}{d(d+1)(d+2)} \delta^{(2)}_{\chi} \, .
\end{equation}
Instead, if we set $\ket{\phi} = \ket{\pmb{0}}$, $||\Pi_{\phi}||_2^2 = d^{-1}$,  we obtain the purity fluctuations of pure stabilizer states
\begin{equation}
\Delta^2 \Pur_A(\ket{\psi}) \big\rvert_{\text{STAB}} = \frac{(d-1)^2}{(d+1)^2(d+2)}  \, .
\end{equation}
Notice that for large $d$, $\Delta^2 \Pur_A(\ket{\psi}) \big\rvert_{\text{STAB}} \sim \Order(d^{-1})$.

\end{appendices}

\end{document}